\documentclass[a4paper,aps,prd,twocolumn,eqsecnum,amsmath,amssymb,nofootinbib,superscriptaddress]{revtex4-2}
\usepackage{dcolumn}
\usepackage{bm}
\usepackage{graphics}
\usepackage{graphicx}
\usepackage{epsfig}
\usepackage{booktabs,multirow}
\usepackage{hhline}
\usepackage{slashed}
\usepackage{rccol}
\usepackage{mathrsfs}
\usepackage{amsmath, amssymb}
\usepackage[dvipsnames]{xcolor,colortbl}
\usepackage{subfigure}
\graphicspath{{figures/}}
\newcommand{\orcid}[1]{\href{https://orcid.org/#1}{\includegraphics[width=10pt]{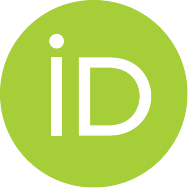}}}

\usepackage{xcolor}\colorlet{linkequation}{blue}
\usepackage[colorlinks=true, 
linkcolor=red,   
filecolor=red,   
citecolor=Green, 
urlcolor=blue,
hyperfootnotes=true]{hyperref}

\newcommand*{\SavedEqref}{}
\let\SavedEqref\eqref
\renewcommand*{\eqref}[1]{%
	\begingroup
	\hypersetup{
		linkcolor=linkequation,
		linkbordercolor=linkequation,
	}%
	\SavedEqref{#1}%
	\endgroup
}

\begin{document}
\title{Constraining neutrino electromagnetic properties with recent low-energy electron recoil data at dark matter direct detection experiments}

\newcommand{\ktu}{Department of Physics,
Karadeniz Technical University, Trabzon, TR61080, Türkiye}

\newcommand{\ktufbe}{Graduate School of Natural and Applied Science, Karadeniz Technical University, Trabzon 61080, Türkiye}

\author{M.~Demirci\orcid{0000-0003-2504-6251}}
\email{mehmetdemirci@ktu.edu.tr}
\affiliation{\ktu}
\author{H.~I.~Sezer\orcid{0009-0005-9349-1930}}
\email{ibrahimhalilsezer@outlook.com}
\affiliation{\ktufbe}
\author{M.~F.~Mustamin\orcid{0000-0003-3996-4651}}
\email{mfmustamin@ktu.edu.tr}
\affiliation{\ktufbe}
\author{ A.~B. Balantekin\orcid{0000-0002-2999-0111}}
\email{baha@physics.wisc.edu}
\affiliation{Department of Physics,
	University of Wisconsin–Madison, Madison, Wisconsin 53706, USA}


\begin{abstract}
Neutrinos that are elastically scattered off atomic electrons provide a unique opportunity to investigate the Standard Model (SM) and beyond SM physics. In this work, we explore the new physics effects of neutrino electromagnetic properties through elastic neutrino-electron scattering using solar neutrinos at the low energy range of PandaX-4T and XENONnT experiments. 
The properties of interest include the neutrino magnetic moment, millicharge, and charge radius, all of which are natural consequences of non-zero neutrino masses.
We investigate their effects by incorporating each property into the Standard Model (SM) framework, given the measured the solar neutrino flux.
By analyzing the latest Run 0 and Run 1 datasets from the PandaX-4T experiment, together with recent results from XENONnT, we derive new constraints on each electromagnetic property of neutrino. We present both flavor-independent results, obtained using a common parameter for all three neutrino flavors, and flavor-dependent results, derived by marginalizing over the three neutrino flavor components.
Bounds we obtained are comparable or improved compared to those reported in the previous studies. 
\end{abstract}

 
\maketitle
\section{Introduction}
The significance of neutrino electromagnetic properties emerged just after Pauli postulated the particle's existence \cite{Pauli:1991}. 
He considered possible magnetic moment neutrinos might possess if they are massive. It was later confirmed by oscillation experiments \cite{KamLAND:2002uet, Super-Kamiokande:2001ljr, SNO:2001kpb} that neutrinos have nonzero mass. 
However, the Standard Model (SM) predicts that neutrinos interact very weakly and are massless, so they have no electric or magnetic properties. Therefore, it is necessary to explore theories beyond the SM (BSM) to explain the existence of massive neutrinos.

Neutrino electromagnetic properties \cite{Bernstein:1963qh, Kayser:1982, Nieves:1982,BahaBalantekin:2018ppj} in various extensions of the SM can be acquired through the effect of quantum loops \cite{Broggini:2012df}. 
This allows neutrinos to interact with electromagnetic fields and charged particles. The implications of these electromagnetic properties can be observed in both astrophysical environments and laboratory experiments \cite{Giunti:2015gga}. In astrophysical environments, neutrinos travel long distances through vacuum and matter, many times encountering magnetic fields. Laboratory experiments employ neutrinos emitted from different sources. Measuring the cross-section for neutrino scattering
on various targets is a rather sensitive and widely used
technique \cite{Ruso:2022qes}. Beyond the neutrino mass, there are other compelling reasons for extending the SM, such as the need to address the possible existence of Dark Matter (DM) and Dark Energy, as well as the matter-antimatter asymmetry in the universe, and various anomalies detected in experiments.

Elastic neutrino-electron scattering \cite{tHooft:1971ucy} offers a powerful tool for exploring the electromagnetic properties of neutrinos. 
This purely leptonic process provides a clean experimental signature for probing the electroweak sector of the SM. A deviation from it also indicates strong evidence of the existence of BSM physics. 
The detection of the process has been a significant achievement in neutrino physics. Since its first detection in high-energy accelerator facilities at CHARM-II \cite{CHARM-II:1992vwk} and E225 \cite{Allen:1992qe}, many experiments measured neutrino -electron elastic scattering including LSND \cite{LSND:2001akn} and TEXONO \cite{TEXONO:2009knm} experiments. 
Additionally, neutrino scattering is used for detecting solar neutrinos in experiments such as Super-Kamiokande \cite{Super-Kamiokande:2002ujc, Super-Kamiokande:2005mbp}, SNO \cite{SNO:2002tuh, SNO:2003bmh}, and BOREXINO \cite{Borexino:2007kvk, Borexino:2008fkj}.

In DM direct detection (DD) experiments, solar neutrinos~\cite{Davis:1968cp, GALLEX:1992gcp, Bahcall:1995mm, Cleveland:1998nv} are capable of inducing both elastic neutrino-electron scattering and coherent neutrino-nucleus scattering processes, producing detectable signals at existing experimental facilities~\cite{Cerdeno:2016sfi, Brdar:2020quo, Schwemberger:2023hee}.
Direct detection of DM has been proposed since the mid-eighties \cite{Drukier:1984vhf, Goodman:1984dc, Drukier:1986tm}. Nowadays, these searches are ongoing in numerous experiments such as XMASS \cite{XMASS:2018bid, XMASS:2020zke, XMASS:2022tkr}, LUX-ZEPLIN (LZ) \cite{LUX:2015abn, LZ:2018qzl,LZ:2023ja},  
DarkSide \cite{DarkSide:2018bpj, DarkSide:2018kuk, DarkSide:2022knj},  XENON \cite{XENON:2020kmp, XENON:2020rca, XENON:2022ltv,XENON:2024wpa}, and PandaX \cite{PandaX:2014mem, PandaX-II:2017hlx, PandaX:2018wtu, PandaX:2022ood, PandaX:2024zbo, PandaX:2024zbo, PandaX:2024med, PandaX:2024cic}. 
For many years, DM direct detection efforts primarily focused on searching for DM-nucleon scattering events.
However, with advancements in detector sensitivity and improved control over low-background environments, electron recoil has emerged as a valuable probe for exploring light DM candidates.
The PandaX-4T \cite{PandaX:2018wtu, PandaX:2024cic, PandaX:2024zbo, PandaX:2024med} and XENONnT \cite{XENON:2022ltv, XENON:2024wpa} experiments have such detectors. The PandaX-4T experiment recently announced robust low-energy electron recoil data \cite{PandaX:2024cic}, using the large liquid xenon (LXe) detector with 1.54 ton$\cdot$year exposure.
The XENONnT experiment, likewise, published low-energy electron recoil data after updating the facility with a large-sized detector; 1.6 ton$\cdot$year exposure, improved background, and reduced systematic uncertainties \cite{XENON:2022ltv}. 
The results from these two experiments have been widely used in several analyses, including weak mixing angle determinations  \cite{Maity:2024aji} and the derivation of limits on different BSM frameworks \cite{A:2022acy, Khan:2023b, Demirci:2025, Giunti:2023yha}.

Motivated by the recent availability of electron recoil data, we study neutrino electromagnetic properties via neutrino-electron scattering, employing solar neutrino flux data from recent direct detection experiments. We analyze the recent electronic recoil data of PandaX-4T \cite{PandaX:2024cic} and XENONnT \cite{XENON:2022ltv}. These experiments reported background signals, one of which is elastic neutrino-electron scattering. We derive new limits on each electromagnetic property from recent PandaX-4T and XENONnT data in the region of electron recoil energy below $30$ keV. We re-analyze the XENONnT data with the consideration of matter effects on the neutrino oscillation as well as radiative corrections to vector couplings, which we discuss in the next section. Our findings confirm previous results from a similar analysis conducted in Ref. \cite{Giunti:2023yha}, and we achieve slightly improved sensitivities from the analysis of XENONnT data.
We further compare our results with existing limits from previous studies available in the literature. These limits are derived from data obtained from artificial neutrino sources, such as reactors, spallation neutron sources, and neutrino beam experiments, as well as from natural neutrino sources.

The remainder of this paper is organized as follows. In Sec.~\ref{sec:nue}, we present the theoretical formulation of the neutrino-electron scattering both in the SM and in the presence of neutrino electromagnetic properties. We also describe the calculation of the differential spectra in this section. In Sec.~\ref{sec:stat}, we outline the data analysis method used for setting limits. In Sec.~\ref{sec:resdis}, we present the expected event spectra for each property and then provide new upper limits from 1 degree of freedom (d.o.f) and 2 d.o.f. analysis and compare them with available limits in the literature. Finally, we conclude our work in Sec.~\ref{sec:summ}.

\section{Theoretical Framework}\label{sec:nue}

\subsection{Neutrino-electron scattering}
The elastic neutrino-electron scattering is a purely leptonic process in the SM. In the process, a neutrino scatters off an electron by the exchange of a charged or neutral boson. 
\begin{figure}[!h]
	\centering
	\includegraphics[scale=0.6]{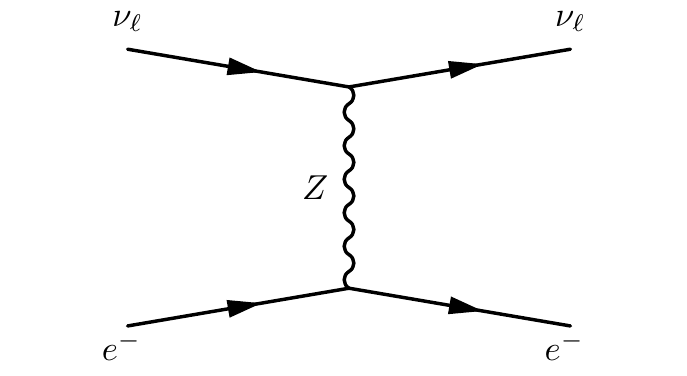}
	\includegraphics[scale=0.6]{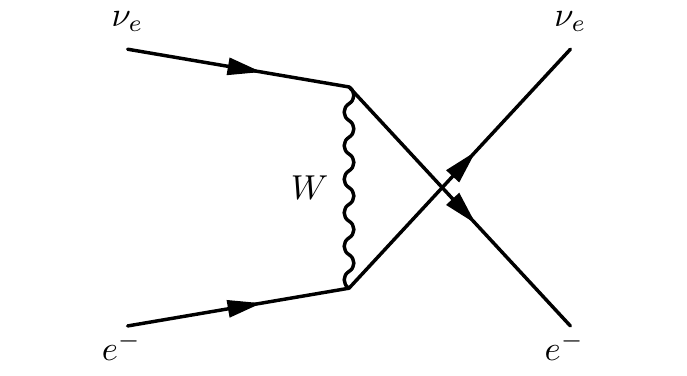}
	\\
	(a) \hspace{35mm} (b)
	\\
	\caption{Feynman diagrams for neutrino-electron scattering via (a) neutral-current (NC) and (b) charge-current (CC) channels in the SM. The index $\ell$ runs for $e,\mu$, and $\tau$ flavor of neutrino.}
	\label{fig:nue_diag}
\end{figure}
The Feynman diagram at tree-level is shown in Fig.~\ref{fig:nue_diag}.
The $Z$ and $W$ represent the neutral and charged boson of the SM, respectively. The differential cross-section for this process in the SM, expressed as a function of the electron recoil energy $T_e$, is given by
\begin{widetext}
	\begin{align} 
		\begin{split}
			\left[\frac{d\sigma}{dT_{e}}\right]_{\mathrm{SM}} = \frac{G_\text{F}^2 m_e}{2\pi} &\Bigg[ (g_V+g_A)^2 + (g_V-g_A)^2 \left(1-\frac{T_{e}}{E_\nu}\right)^2 - (g_V^2-g_A^2) \frac{m_e T_{e}}{E_\nu^2} \Bigg],
		\end{split}
		\label{eq:sm_nue}
	\end{align}
\end{widetext}
where $m_e$ electron mass, $E_\nu$ is the initial neutrino energy, and 
$G_\text{F}$ is the Fermi coupling constant.
Note that for the anti-neutrino case, the relative sign between $g_V$ and $g_A$ at the first and second term is reversed. In the SM, the vector coupling is
\begin{align}
	g_V =& -\frac{1}{2}+2\sin^2\theta_W + \delta_{\ell e},
\end{align}
while the axial-vector coupling is
\begin{align}
	g_A =& -\frac{1}{2}+\delta_{\ell e},
\end{align}
where the Kronecker symbol $\delta_{\ell e}$ stands for NC and CC nature of the process. The weak mixing angle $\sin^2\theta_W$ is given as $\sin^2\theta_W= 0.23873$ in the $\overline{\text{MS}}$ scheme at the low-energy regime \cite{ParticleDataGroup:2024cfk}. Both couplings depend on the flavor $\ell=e,\mu,\tau$ of the incoming neutrino.
We consider favor-dependent couplings as $g_V^{\nu_e e}=0.9524, g_A^{\nu_e e}=0.4938, g_V^{\nu_\mu e}=-0.0394, g_A^{\nu_\mu e}=-0.5062, g_V^{\nu_\tau e}=-0.0350$, and $g_A^{\nu_\tau e}=-0.5062$ when taking into account radiative corrections \cite{Erler:2013c, AtzoriCorona:2022jeb} and the
latest weak mixing angle calculation \cite{AtzoriCorona:2025ygn}.

A notable feature of neutrino-electron scatterings is that the process is highly directional, i.e. the outgoing electron tends to have very small angles relative to the direction of the incoming neutrino  \cite{Formaggio:2012cpf}. This property has been leveraged in several experiments involving neutrinos, especially for detecting solar neutrinos, such as Super-Kamiokande \cite{Super-Kamiokande:2002ujc, Super-Kamiokande:2005mbp}, SNO \cite{SNO:2002tuh, SNO:2003bmh}, and BOREXINO \cite{Borexino:2007kvk, Borexino:2008fkj} experiments.
Neutrinos originating from the Sun can induce elastic neutrino-electron scattering events in DD experiments. Both neutrino-electron scatterings and coherent elastic neutrino-nucleus scatterings serve as significant backgrounds in these experiments. Although the signal is expected to be small, it remains challenging to eliminate it completely in practice. 

\subsection{Neutrino electromagnetic properties}
A precise understanding of the intrinsic properties of neutrinos is crucial for exploring how the SM should be expanded in the lepton sector. Neutrinos are neutral particles within the SM, which means they cannot couple directly with photons at the tree level. 
However, despite having no electric charge, neutrinos are massive particles and possess an electromagnetic structure due to loop level corrections. This includes properties such as the neutrino magnetic moment ($\nu$MM), millicharge ($\nu$MC), and charge radius ($\nu$CR). These characteristics emerge at the loop level in SM or beyond the SM, enabling neutrinos to engage in electromagnetic interactions with photons and charged particles. In the one-photon approximation, these properties can be combined into an effective electromagnetic current \cite{Kayser:1982, Nieves:1982}.
\begin{align}
	\langle \nu_{f} | j_{\text{EM}}^\mu | \nu_{i} \rangle = \bar{u}_{f} \Lambda^\mu_{fi} (\text{q}) u_{i},
\end{align}
where the photon four-momentum transfer is represented by $\text{q}$, whereas the initial and final neutrino spinor are given by $u_{i}$ and $\bar{u}_{f}$, respectively. The vertex function $\Lambda_{fi}^\mu$ denotes a $3\times 3$ matrix in the neutrino mass eigenstate that encodes the neutrino electromagnetic properties. For low-energy case, $\text{q}\rightarrow 0$, this vertex can be written as
\begin{align}
	\Lambda_{{fi}}^\mu (\text{q}) = \left( q_{\nu_{fi}}-\frac{\text{q}^2}{6}\langle r^2_{\nu_{fi}} \rangle \right) \gamma^\mu - i\sigma^{\mu\rho} \text{q}_\rho \mu_{\nu_{fi}}.
\end{align}
In the case where $f=i$, we obtain the diagonal or effective terms, while for $f\neq i$, we analyze the transition neutrino electromagnetic properties.

In this work, we focus on the effective case of neutrino electromagnetic properties: neutrino magnetic moment $\mu_{\nu}$, millicharge $ q_{\nu}$ and charge radius $\langle r^2_{\nu} \rangle$. "Effective" here indicates the potential inclusion of contributions from electric and anapole moments to the magnetic moment and charge radius, respectively.

\subsubsection{Neutrino magnetic moment ($\nu$MM)}
The neutrino magnetic moment is one of the most extensively studied electromagnetic properties because it can arise in various models that involve massive neutrinos. Specifically, in the minimal extension of the SM, where neutrinos acquire Dirac masses via the introduction of right-handed neutrinos, the $\nu$MM is expressed as \cite{Giunti:2015q}.
\begin{align}
  \mu_\nu = \frac{3 e_0 G_\text{F}}{8 \sqrt{2} \pi^2}m_\nu \simeq 3.2 \times 10^{-19} \frac{m_\nu}{\text{eV}} \mu_\mathrm{B},
\end{align}
where $m_\nu$ is the neutrino mass, $e_0$ is the electric charge and  $\mu_\mathrm{B}$ is Bohr magneton. Given the current upper limit on neutrino mass, this value is less than $\mu_\nu \sim 10^{-18} \mu_\mathrm{B}$, which is too small to be observed experimentally \cite{Balantekin:2013sda}. However, given that the $\nu$MM is predicted to be larger in some BSM scenarios, a positive observation would be a clear physics signal beyond the minimally extended SM.

The $\nu$MM's contribution can be calculated by adding incoherently to the neutrino-electron scattering in the SM since their interaction flips the helicity of ultra-relativistic neutrinos. Note that in the SM, weak interaction is helicity-conserving. Its effect, therefore, can be studied by adding the $\nu$MM cross-section \cite{Vogel:1989q}
\begin{align}
    \left[\frac{d\sigma}{dT_e}\right]_{\nu\mathrm{MM}} = \frac{\pi \alpha_{\mathrm{EM}}^2}{m_e^2}\left(\frac{1}{T_e}-\frac{1}{E_\nu}\right)\left|\frac{\mu_{\nu_\ell}}{\mu_\mathrm{B}}\right|^2,
\end{align}
where $\alpha_{\mathrm{EM}}$ is the fine-structure constant, and $\mu_{\nu_\ell}$ is an effective neutrino MM. For our case of low electron recoil energy, $T_e$, this cross-section is proportional to the inverse of this recoil energy. Therefore, the effects of the $\nu$MMs are probed by the observation of an event excess near the $T_e$ threshold. 

\subsubsection{Neutrino millicharge ($\nu$MC)}
The electric neutrality of neutrinos in the SM is guaranteed by charge quantization \cite{Geng:1989q, Babu:1990q}. However, in some BSMs, they can have a very small electric charge (usually referred to as millicharge) (see Refs. \cite{Giunti:2015q, Das:2020q} and references therein), enabling it to couple to the photon. 

The $\nu$MC contribution to the SM cross-section of neutrino-electron scattering can be obtained by substituting
\begin{align}
    \left[\frac{d\sigma}{dT_e}\right]_{\mathrm{SM}+\nu\mathrm{MC}} = \left[\frac{d\sigma}{dT_e}\right]_{\mathrm{SM}} \biggl(g_V \rightarrow g_V - \frac{\sqrt{2} \pi \alpha_{\text{EM}}}{G_\text{F} m_e T_e} q_{\nu_\ell} \biggr).
\end{align}
%
The neutrino-electron scattering process provides full information on the charge $q_{\nu_\ell}$ of the neutrino flavor. 
The effect of this charge is greater at low electron recoil values, which allows for the potential observation of very small electric charges in low-threshold facilities.

\subsubsection{Neutrino charge radius ($\nu$CR)}
Practically, the charge radius of a neutrino is generated by a loop insertion into the $\nu_\ell$ line, where the charged lepton(s) $\ell$ and $W$ boson(s) can enter. In SM, these contributions can be calculated as \cite{Bernabeu:2000,Bernabeu:2004}
\begin{align}
    \langle r_{\nu_\ell}^2\rangle_{\mathrm{SM}}=-\frac{G_\text{F}}{2\sqrt{2}\pi^2}\left[3-2\ln({m_\ell^2/m_W^2})\right]
\end{align}
with $m_W$ and $m_\ell$ are the $W$-boson and charged lepton masses, respectively.
Numerical values of the $\nu$CR are predicted by SM as \cite{Bernabeu:2004}
\begin{eqnarray}
	\begin{split}
    \left\langle r_{\nu_e}^2\right\rangle_{\text{SM}}&=-8.3\times{10}^{-33}\text{cm}^2
    \\
    \left\langle r_{\nu_\mu}^2\right\rangle_{\text{SM}}&=-4.8\times{10}^{-33}\text{cm}^2
    \\
    \left\langle r_{\nu_\tau}^2\right\rangle_{\text{SM}}&=-3.0\times{10}^{-33}\text{cm}^2
    \end{split} \label{eq:SMvalues}
\end{eqnarray}
for each flavor, respectively. 
It is worth noting that these are the only charge radii that exist in the SM where the neutrino flavor is conserved. The $\nu$CR affects the scattering of neutrinos with charged particles. Therefore, it contributes to the neutrino-electron scattering process. This contribution can be obtained by substituting into the SM cross-section as follows
\begin{align}
    \left[\frac{d\sigma}{dT_e}\right]_{\mathrm{SM}+\nu\mathrm{CR}} = \left[\frac{d\sigma}{dT_e}\right]_{\mathrm{SM}} \biggl(g_V \rightarrow g_V + \frac{\sqrt{2} \pi \alpha_{\text{EM}}}{3 G_\text{F} } \langle r_{\nu_\ell}^2\rangle \biggr).
\end{align}
This term fully encodes the contribution of neutrino charge radii $\nu$CRs.
We note that there might be a momentum dependence, and hence electronic recoil, which comes from radiative corrections of the neutrino charge radius \cite{AtzoriCorona:2024rtv}. However, the impacts are relevant for momenta larger than the mass of the corresponding charged lepton within the loops. This is expected to be negligible for the typical momenta of direct-detection experiments considered in this work.

\subsection{Event Rate Spectra} \label{sec:ER}
The differential event rate, obtained by multiplying the differential cross section by the incident neutrino flux, can be expressed as
\begin{align}
	\left[\frac{dR}{dT_{e}}\right]^i_X = Z_{\text{eff}}(T_e) \int_{E_{\nu}^{\text{min}}}^{E_{\nu,i}^{\text{max}}} dE_\nu \frac{d\Phi^i}{dE_\nu} \left[\frac{d\sigma}{dT_{e}}\right]^{\nu e}_X,
\end{align}
where the index $X=\mathrm{SM}, \nu\mathrm{MM}, \nu\mathrm{CR}, \cdots$ denotes different contributions. 
The minimum neutrino energy $E_{\nu}^{\text{min}}$ required to produce an electron recoil energy is
\begin{align}
	E_{\nu}^{\text{min}} = \frac{1}{2}\left(T_e+\sqrt{T_e^{2} + 2m_e T_e} \right),
\end{align}
while the upper limit, $E_{\nu,i}^{\text{max}}$, corresponds to the endpoint of each neutrino flux spectrum.
The solar neutrino flux $d\Phi^i(E_\nu)/dE_\nu$ represents individual flux components, labeled by $i$. In this study, only the $pp$ and $^7$Be fluxes are considered, as they provide the dominant contributions to the electron recoil event rates in the low-energy region probed by direct detection experiments. 

As the electron recoil $T_e$ approaches a few keV, which is comparable to the electron binding energy, the approximation of treating electrons as free is generally no longer valid due to the effects of atomic ionization \cite{Chen:2016eab}. To address this, we need to multiply the neutrino-electron scattering cross-section by the number of effective electron charges that can be ionized. This is represented by
\begin{align}
Z_{\mathrm{eff}}(T_e)=\sum_{\alpha}n_\alpha\theta(T_e - B_\alpha)
\end{align}
where $n_\alpha$ denotes the number of electrons and $B_\alpha$ represents their binding energy in atomic shell $\alpha$. The function $\theta(x)$ refers to a Heaviside step function. We take their values for the xenon target from Ref. \cite{xraydata:2009}.
This is needed to correct the cross-section computed within the free electron approximation, which neglects atomic binding effects by treating electrons as free and at rest \cite{Kouzakov:2017hbc, Hsieh:2019hug}.
Neutrinos undergo flavor oscillations as they propagate from the Sun to the Earth. As a result, solar neutrinos arrive at the detector as a mixture of $\nu_e$, $\nu_\mu$, and $\nu_\tau$. 
In order to account for this effect, the cross-section must be weighted by the corresponding survival and transition probabilities
\begin{align}
	\left[\frac{d\sigma}{dT_{e}}\right]^{\nu e}_X = P_{ee} \left[\frac{d\sigma_{\nu_e}}{dT_{e}}\right]_X + \sum_{f=\mu,\tau} P_{ef} \left[\frac{d\sigma_{\nu_f}}{dT_{e}}\right]_X,
\end{align}
where the conversion probabilities of $\nu_e$ to $\nu_\mu$ and $\nu_\tau$ are given by $P_{e\mu}=(1-P_{ee})\cos^2\vartheta_{23}$ and $P_{e\tau}=(1-P_{ee})\sin^2\vartheta_{23}$. The $P_{ee}$ denotes the survival probability of $\nu_e$ which satisfies \cite{Maltoni:2015kca}
\begin{align}
	\begin{split}
		P_{ee} = & \cos^2 (\vartheta_{13}) {\cos^2 (\vartheta_{13}^m)}\left( \frac{1}{2} + \frac{1}{2} \cos (2\vartheta_{12}^m) \cos (2\vartheta_{12}) \right) \\ &+ \sin^2 (\vartheta_{13}) {\sin^2(\vartheta_{13}^m)}.
		\label{Pee}
	\end{split}
\end{align}
where the label $m$ denotes the matter effect contribution. The survival probabilities depend on neutrino mixing angles of $\vartheta_{12}, \vartheta_{13}$ and $ \vartheta_{23}$.
In this work, we include the day-night asymmetry effect, which arises due to the interaction of neutrinos with the Earth's matter, in the evaluation of these probabilities. 
The normal-ordering neutrino oscillation parameters are adopted from the latest global 3-flavor analysis provided by NuFit-5.3, excluding the Super-Kamiokande atmospheric neutrino data set \cite{Esteban:2020cvm}. It is important to note that the transition probabilities allow us to separate the neutrino flavor that arrives at the detector. It is then feasible to constrain separately each electromagnetic property of the neutrino according to its flavor.

The predicted number of elastic neutrino-electron scattering events is calculated from
\begin{align}
	\begin{split}
		R^k_X=&\varepsilon N_{t}\int_{T_e^k}^{T_e^{k+1}}dT_e \text{ } \mathcal{A}(T_e) \int_{0}^{T_e^{'\text{max}}}dT_e' \text{ } \mathcal{R}(T_e,T_e') \\ &\times \sum_{i=pp,^7\text{Be}} \left[\frac{dR}{dT'_{e}}\right]^i_X,
	\end{split}
\end{align}
where the factor $N_{t}$ 
represents the number of target nuclei per unit mass of the detector material. 
The variables $T_e$ and $T_e'$ denote the reconstructed and true electron recoil energies, respectively. 
The functions $\mathcal{A}(T_e)$ and $\mathcal{R}(T_e,T_e')$ 
correspond to the detector efficiency and the energy resolution (smearing) function, both of which are essential for accurately modeling the observed signal. Detector efficiencies for PandaX-4T and XENONnT are taken from Refs.~\cite{PandaX:2024cic} and \cite{XENON:2022ltv}, respectively. We use the normalized Gaussian smearing function with energy resolution $\sigma =  0.073+0.173 T_e - 0.0065 T_e^2  + 0.00011 T_e^3$ \cite{PandaX:2022ood} and $\sigma=0.31\sqrt{T_e} + 0.0037 T_e$ \cite{XENON:2020rca} for PandaX-4T and XENONnT, respectively. 
To obtain the event count reported by the experiments, the differential event rate is further multiplied by the exposure factor $\varepsilon$.  The Run0 dataset of the PandaX-4T experiment corresponds to an exposure of 
$198.9\text{ ton} \cdot \text{day}$, while the Run1 dataset has an exposure of $363.3\text{ ton} \cdot \text{day}$. In comparison, the exposure of the XENONnT SR0 dataset amounts to 
$1.16 \text{ ton}\cdot\text{ year}$.  
Lastly, the maximum recoil energy is determined by the kinematic constraints of the neutrino-electron scattering process and satisfies
\begin{align}
	T_e^{'\text{max}} = \frac{2E_\nu^2}{2E_\nu + m_e}.
\end{align}
This relation explains that lighter targets improve the maximum recoil energy produced in the detector.

\section{Data Analysis Details}\label{sec:stat}
We now present the implementation of our statistical analysis using the latest experimental data. In particular, we analyze recent low-energy electron recoil measurements reported by the PandaX-4T~\cite{PandaX:2024cic} and XENONnT~\cite{XENON:2022ltv} collaborations.
The PandaX-4T recently published their unblinding Run0 and Run1 data in Ref. \cite{PandaX:2024zbo} with electron recoil $<30$ keV. The PandaX-4T experiment is conducted at the China Jinping Underground Laboratory and employs a time projection chamber (TPC) containing 3.7 tonnes of liquid xenon (LXe) to accurately determine the position, timing, and energy of individual events~\cite{PandaX:2018wtu}. 
The Run0 data was obtained after operating from 28 November 2020 to 16 April 2021, while the Run1 data was obtained from 16 November 2021 to 15 May 2022.
The XENONnT is the latest detector of the XENON program, located at the Laboratori Nazionali del Gran Sasso, featuring a dual-phase xenon TPC with 8.5 tons of LXe. The electron recoil SR0 dataset, collected from 6 July 2021 to 10 November 2021, was published by the experiment with no excess above the background \cite{XENON:2022ltv}.
Apart from the main objective of observing DM candidates, data from both experiments have been utilized to probe some new physics searches, such as light mediator models, dark photon, and also neutrino electromagnetic properties. 
The neutrino-electron scattering, as one of the background sources in these facilities, contributes a well-known flat signal. This implies that the signal can be subtracted in standard DM analyses 
In addition, BSM signatures in this channel could influence the observation signal, making it a potential framework to study the effect of neutrino electromagnetic properties.
Accordingly, we analyze the contribution of each neutrino's electromagnetic property in the region of interest (ROI) of both PandaX-4T and XENONnT experiments.

For statistical analysis of the new physics parameter(s) of interest $\mathcal{S}$, we employ a Poissonian $\chi^2$ function \cite{Baker:1983tu, Fogli:2002pt} as follows 
\begin{widetext}
	\begin{align}
		\begin{split}
			\chi^2(\mathcal{S}) = \mathrm{min}_{(\alpha_i,\beta_j)} 2  \sum_{k=1}^{30} \biggl[  R_\text{exp}^{k}(\mathcal{S}; \alpha,\beta) - R_{obs}^{k} + R_{obs}^{k} \ln \Bigg( \frac{R_{obs}^{k} }{R_\text{exp}^{k}(\mathcal{S}; \alpha,\beta) }\Bigg)  \biggr]  + \sum_{i}\left(\frac{\alpha_i}{\sigma_{\alpha_i}}\right)^2 +  \sum_{j}\left(\frac{\beta_j}{\sigma_{\beta_j}}\right)^2,
		\end{split} \label{eq:chi2}
	\end{align} 
\end{widetext}
where the $R_{\text{obs}}^k$ and $R_{\text{exp}}^k$ represent the observed and expected event rates in the $k$-th energy bin, respectively. The expected one consists of SM $R^k_{\text{SM}}$ plus BSM $R^k_{\text{BSM}}$ ($\nu\mathrm{MM}, \nu\mathrm{MC}, \nu\mathrm{CR}, \cdots$) contribution and other background components $R_{\text{Bkg}}$. 
Explicitly, we adopt the form $R_{\text{exp}}^k(\mathcal{S};\alpha_i, \beta_j)=(1+\alpha_i)(R^k_{\text{SM}}+R^k_{\text{BSM}}(\mathcal{S})) + (1+\beta_j)R_{\text{Bkg},j}^k$
where the parameters $\alpha_i$ and $\beta_j$ are uncertainty coefficients associated with the solar neutrino fluxes and normalization factors applied to each individual background component, respectively. 
We use the data presented in Fig. 2 of Ref. \cite{PandaX:2024cic} for PandaX-4T and the data in Fig. 4 (5) of Ref. \cite{XENON:2022ltv} for XENONnT with recoil energies below 30 keV.
The experimental uncertainty for each energy bin is denoted by $\sigma^k$.
The $\sigma_{\alpha}$ denotes the uncertainty of solar neutrino flux, as listed in Table 6 of Ref.~\cite{Vinyoles:2016djt}, while $\sigma_{\beta}$ denotes the experimental background uncertainty.
The total event rates, including both SM and new physics contributions, are computed using the solar neutrino fluxes based on Bahcall’s spectrum \cite{Bahcall:1989ks}, normalized according to the B16-GS98 Standard Solar Model \cite{Vinyoles:2016djt}.
We note that the total neutrino-electron scattering event rates obtained by integrating over all energy bins using this setup are in good agreement with the results reported by the collaborations. The predicted event counts are 41.65 and 74.68 for PandaX-4T Run0 and Run1, respectively \cite{PandaX:2024cic}, and 303.48 for XENONnT up to 140 keV \cite{XENON:2022ltv}, consistent with experimental results. In the low-energy region of interest, the corresponding value for XENONnT is 77.95.
\begin{figure*}[htb]
	\centering
	\includegraphics[scale=0.38]{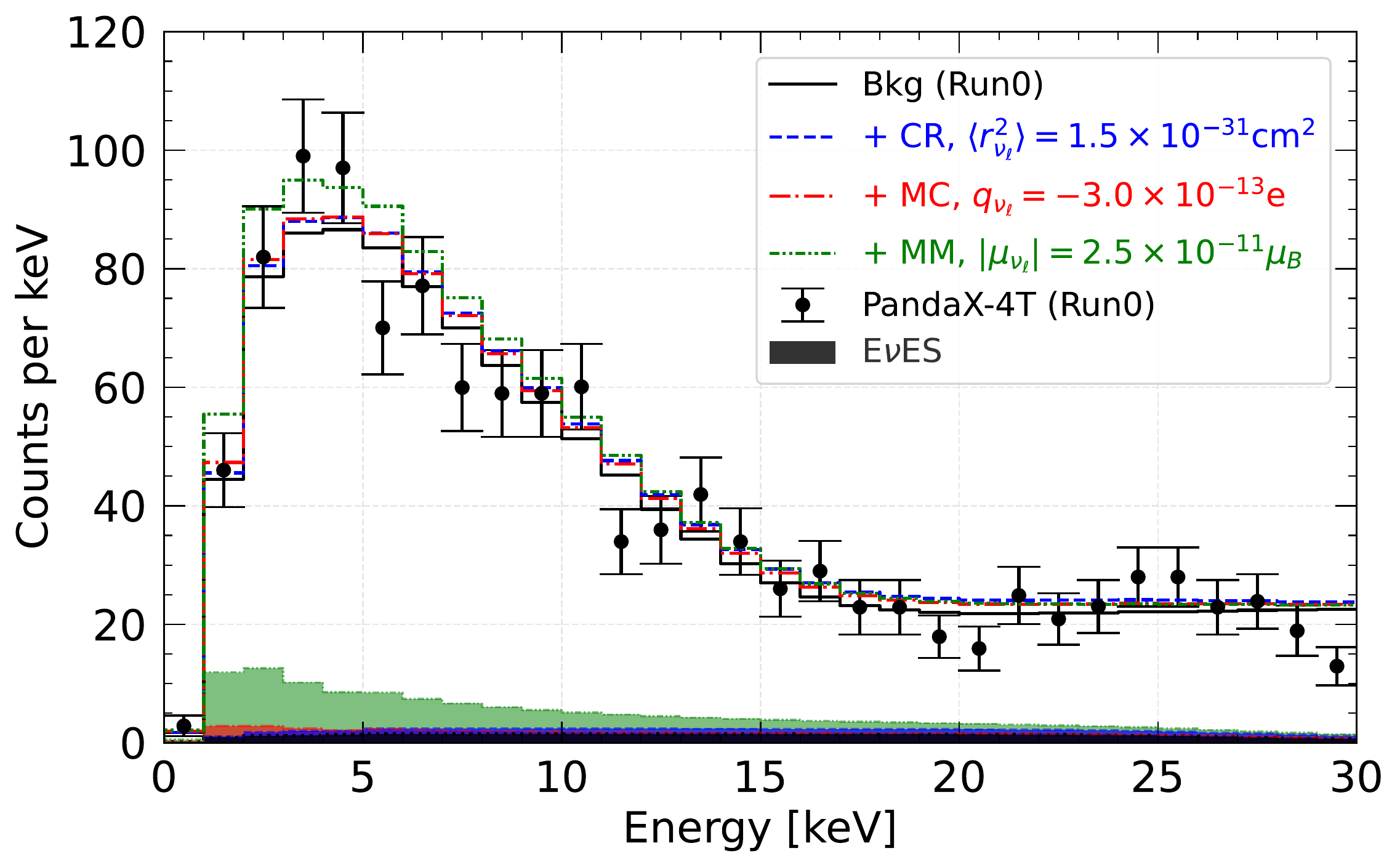}
	\includegraphics[scale=0.38]{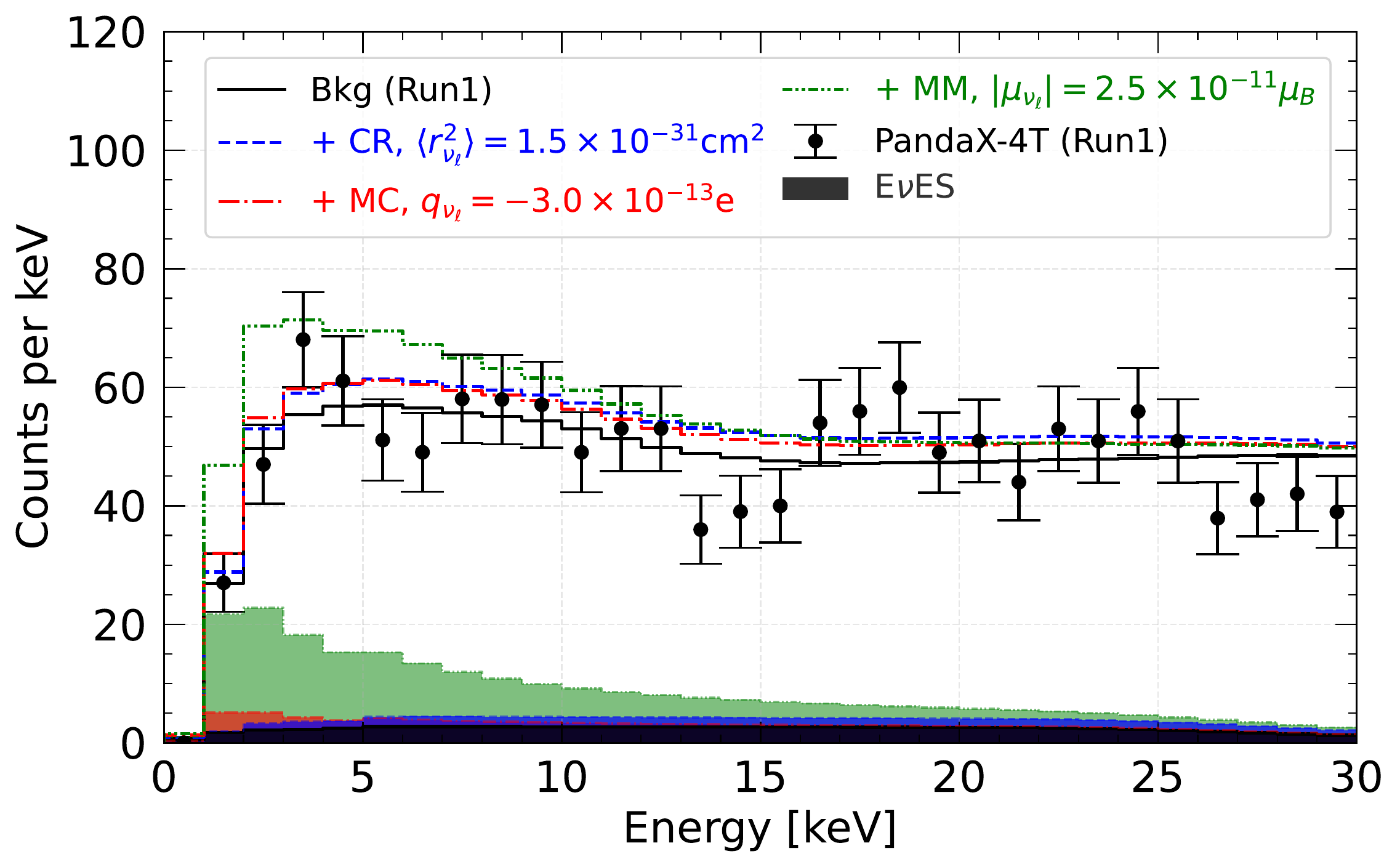}
	\\
	\hspace{8mm} (a) \hspace{84mm} (b)
	\\
	\includegraphics[scale=0.38]{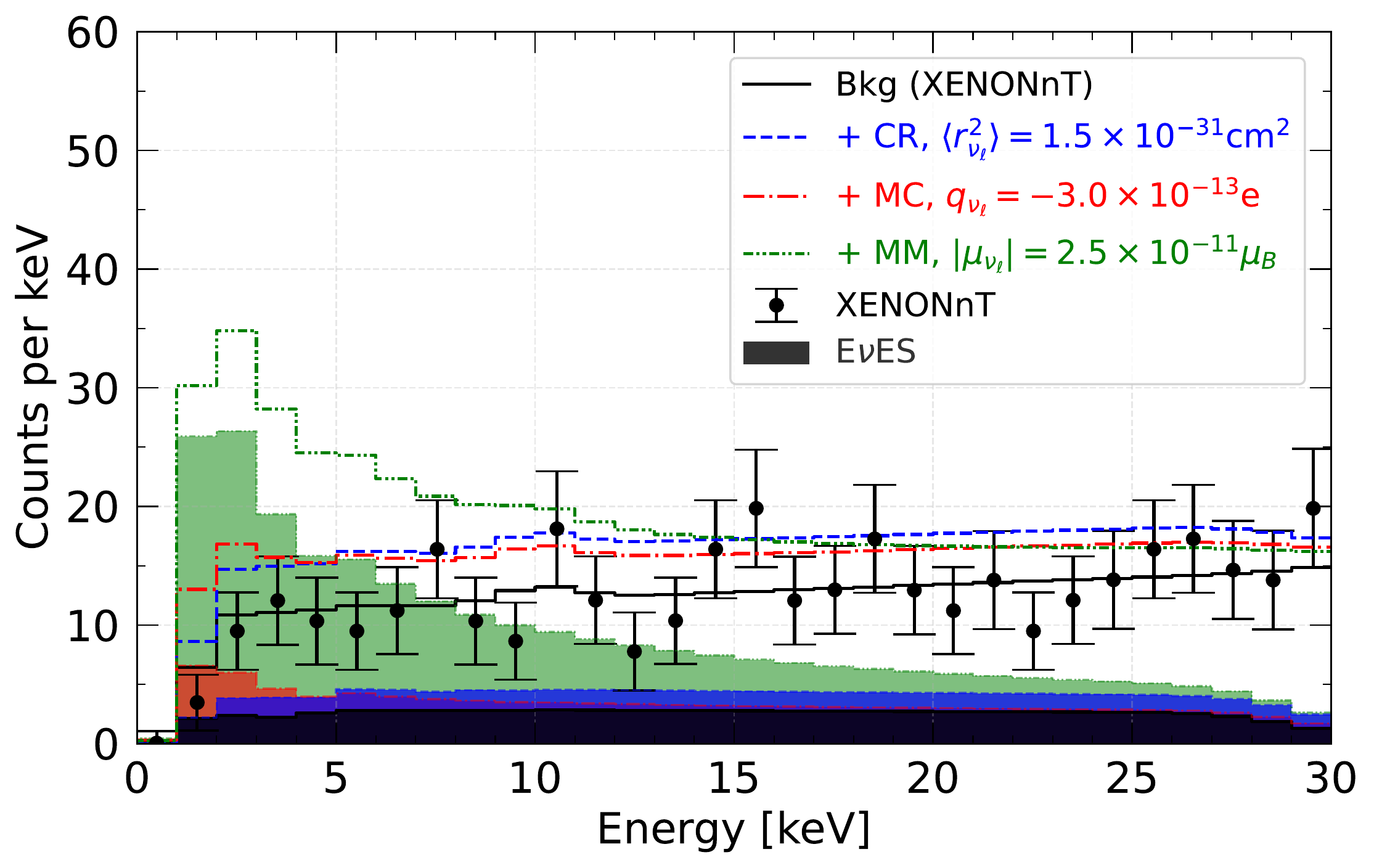}
	\\
	\hspace{6mm} (c)
	\\
	\vspace{-2mm}
	\caption{Predicted signals of each neutrino electromagnetic property as a function of electron recoil energy for the PandaX-4T (a) Run0 and (b) Run1, and (c) XENONnT. 
	Individual contributions to the E$\nu$ES are shown as filled colored histograms, while their contributions to the total background are shown as empty, outlined histograms in the same colors.}
	\label{fig:rate_bsm}
\end{figure*}
\section{Results and Discussion}\label{sec:resdis}

We present numerical results for the neutrino electromagnetic properties from the analysis of DD experiments data in this section. We first discuss the expected event spectra and then current bounds on neutrino magnetic moments, millicharge, and charge radius.
\begin{figure*}[htb]
	\centering
	\includegraphics[scale=0.42]{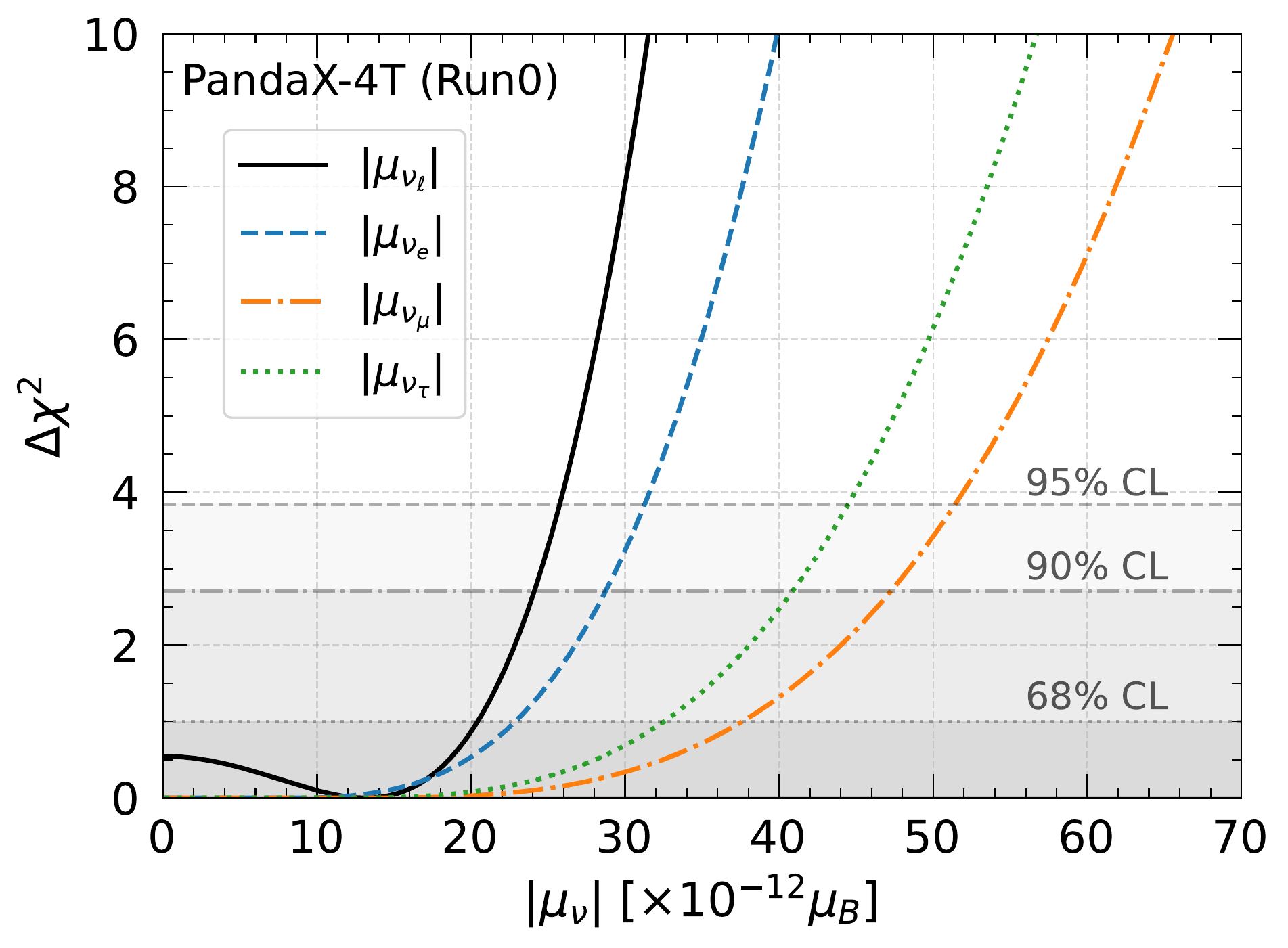}
	\includegraphics[scale=0.42]{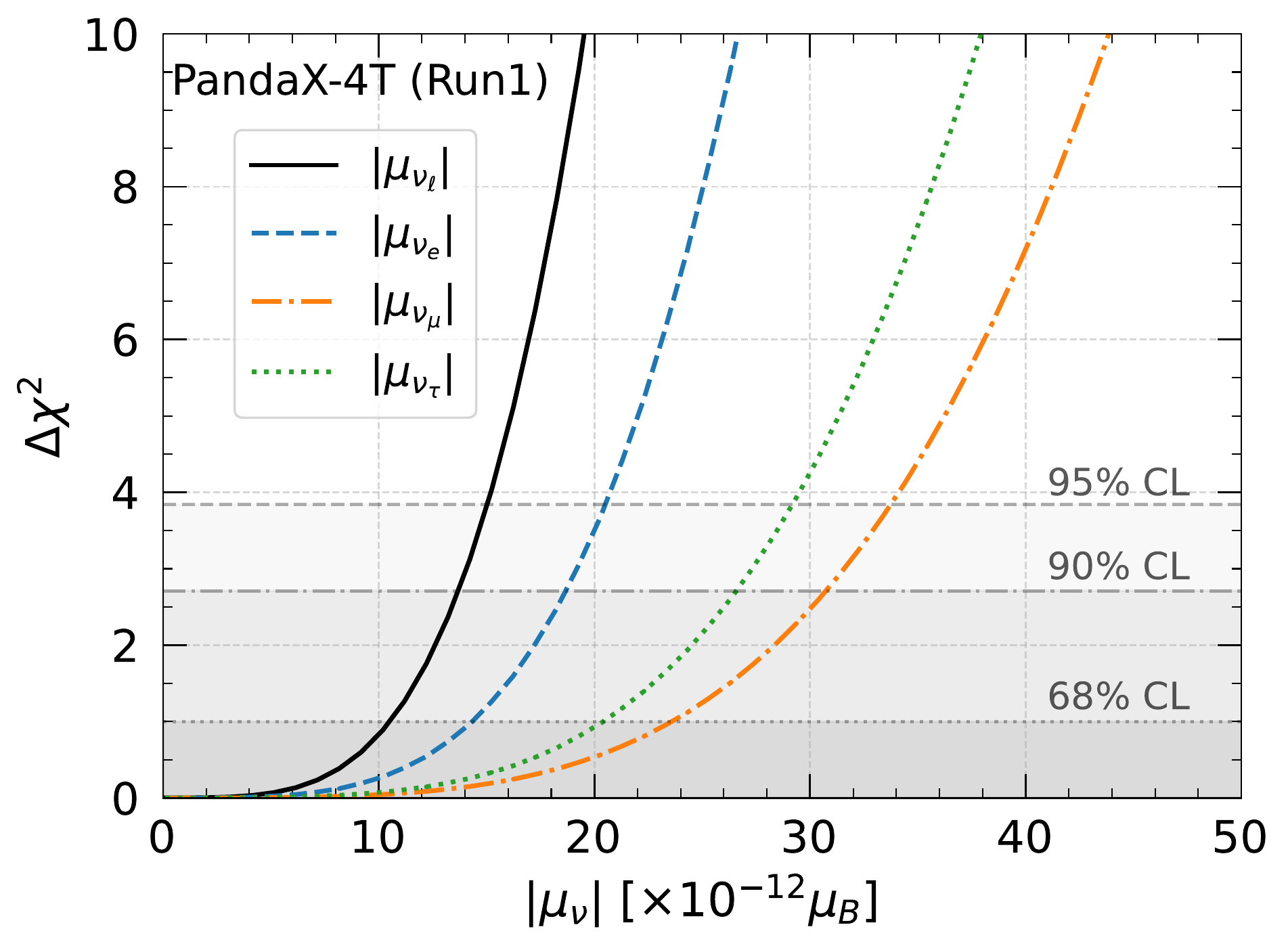}
	\\
	\hspace{7.0mm} (a) \hspace{77mm} (b)
	\\
		\includegraphics[scale=0.42]{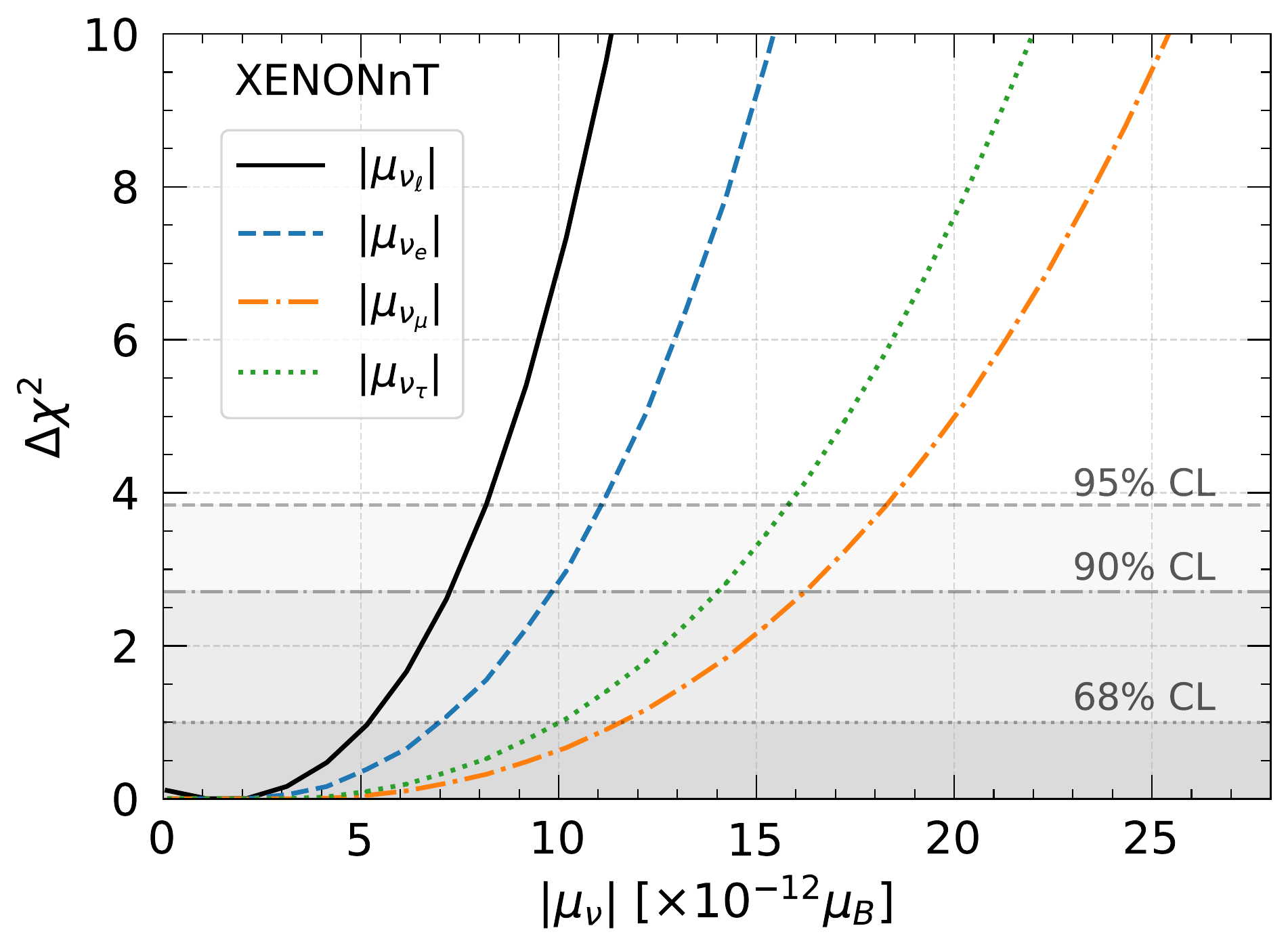}
        \\
		\hspace{8.8mm} (c)
	\vspace{-2mm}
	\caption{One-dimensional $\Delta\chi^2$ distributions of neutrino magnetic moment from the analysis of recent data: (a) PandaX-4T Run0, (b) PandaX-4T Run1, and (c) XENONnT. Results are presented for both the flavor-independent effective neutrino magnetic moment and the flavor-dependent cases obtained via marginalization over the three neutrino flavors.}
	
	\label{fig:nuMM}
\end{figure*}
\begin{figure*}[htb!]
	\centering
	\includegraphics[scale=0.315]{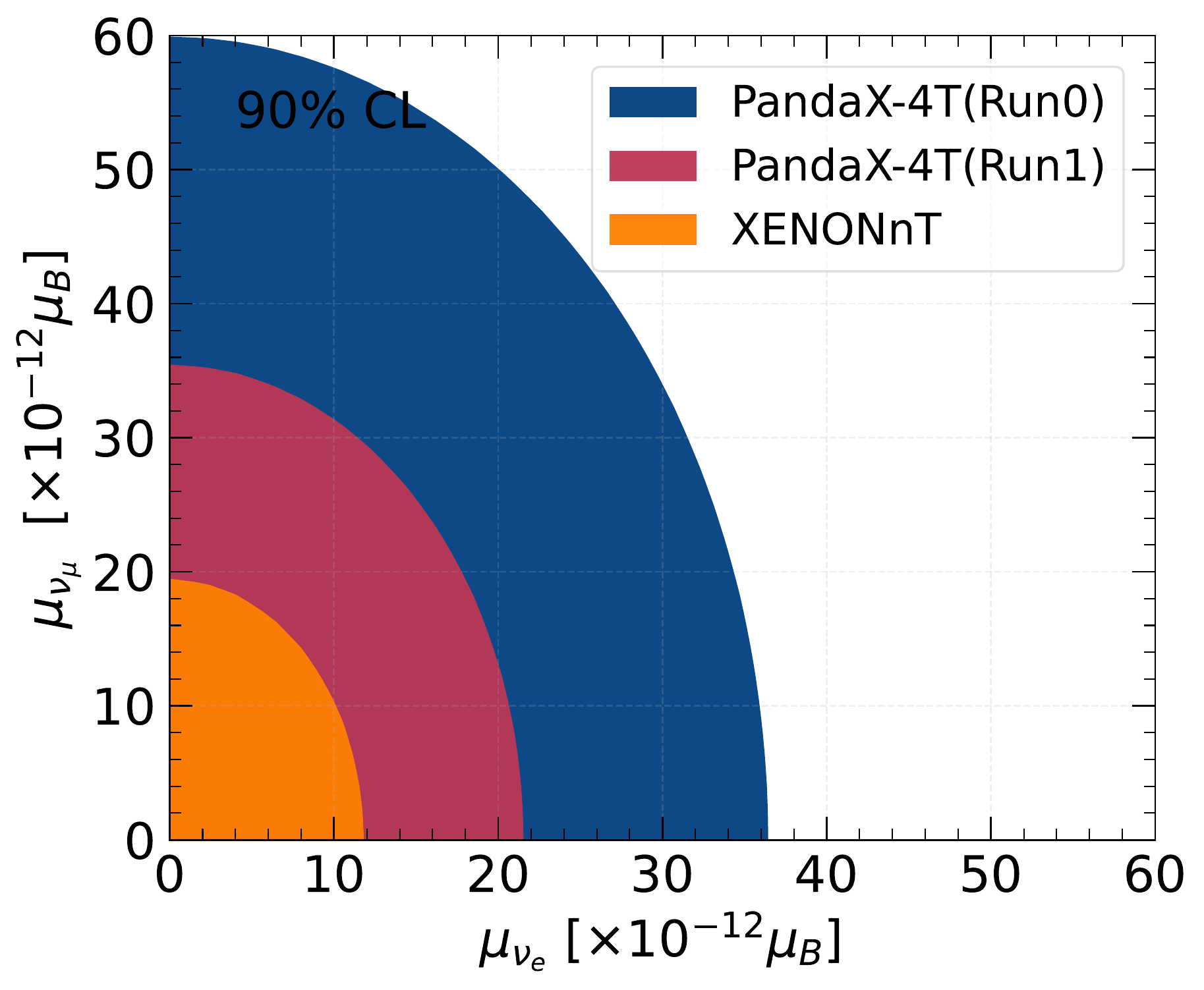}
	\includegraphics[scale=0.315]{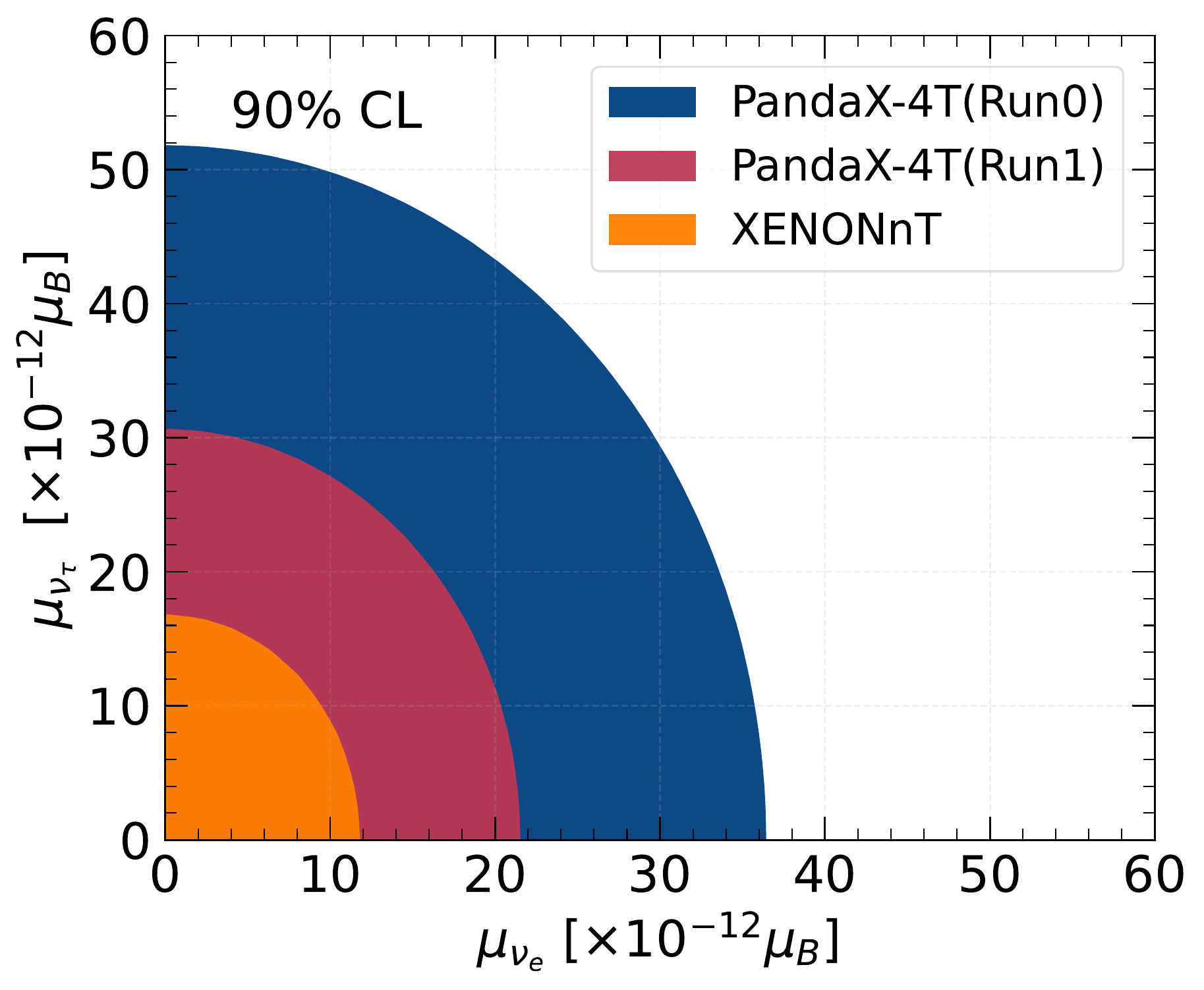}
    \includegraphics[scale=0.315]{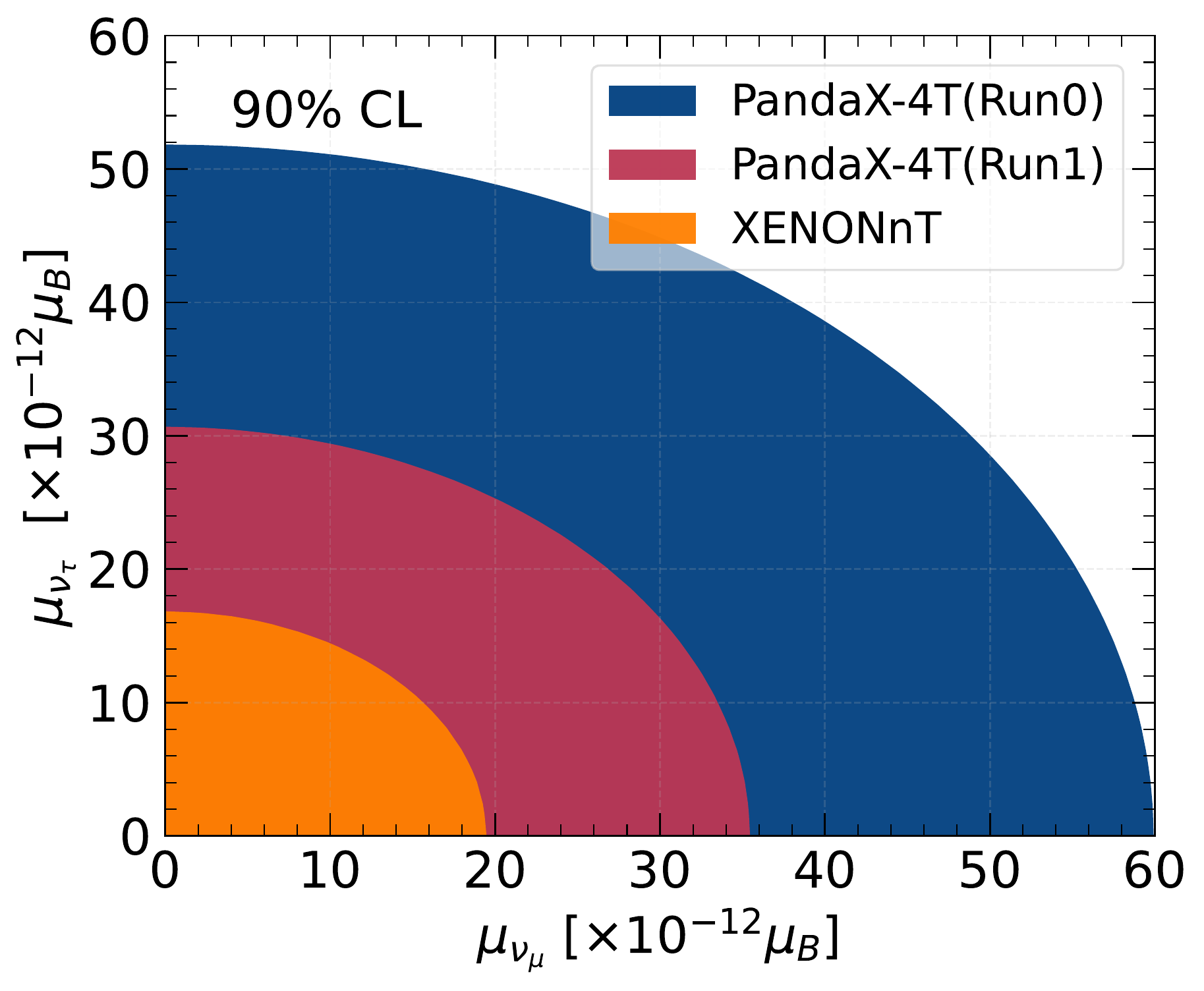}
	\\
	\hspace{7mm} (a) \hspace{53mm} (b) \hspace{53mm} (c)
	\vspace{-2mm}
	\caption{The $90\%$ CL allowed regions with 2 d.o.f. in the (a) $(\mu_{\nu_e},\mu_{\nu_\mu})$, (b)  $(\mu_{\nu_e},\mu_{\nu_\tau})$ and (c)  $(\mu_{\nu_\mu},\mu_{\nu_\tau})$ planes, obtained by marginalizing over the remaining neutrino flavor components.
    The results correspond to the analysis of PandaX-4T-Run0 (blue region), PandaX-4T-Run1 (red region), and  XENONnT data (orange region). 
	}
	\label{fig:nuMM_2dof}
\end{figure*}

In Fig. \ref{fig:rate_bsm}, we compare the electron recoil data of PandaX-4T and XENONnT with the expected signal in the presence of neutrino electromagnetic properties. We take the contribution of the total background from each experiment and subtract our predicted solar neutrino component from it. The signals are shown as dashed-black lines, which are in good agreement with the published results of the corresponding experiments. We set benchmarks for each electromagnetic property of the neutrino and incorporate them with the SM prediction. The contribution of each neutrino's electromagnetic property to the E$\nu$ES process is shown as filled colored histograms, while to the total background as empty, outlined histograms in the same colors. 
We can see significant deviations in the low recoil energies, especially for the magnetic moment and millicharge, as we consider the existence of these BSM scenarios, pointing out the importance of advancing our technology to search in this territory.

For each electromagnetic property, we derive the 1 d.o.f. limits by following the $\chi^2$-analysis, discussed in the previous section. 
Here, the 68\%, 90\%, and 95\% confidence level (CL) limits for each quantity are marked.  We present both the flavor-independent results, which are calculated by taking a common parameter for the three neutrino fluxes, and flavor-dependent results. In the flavor-independent case, the dependence on the oscillation parameters cancels out because we have the same value for all flavors.
Furthermore, we provide the 90\% CL allowed regions by 2 d.o.f. analysis. 
In the flavor-dependent cases, the results are obtained by marginalizing over the three neutrino flavor components. 

\begin{figure*}[htb]
	\centering
	\includegraphics[scale=0.42]{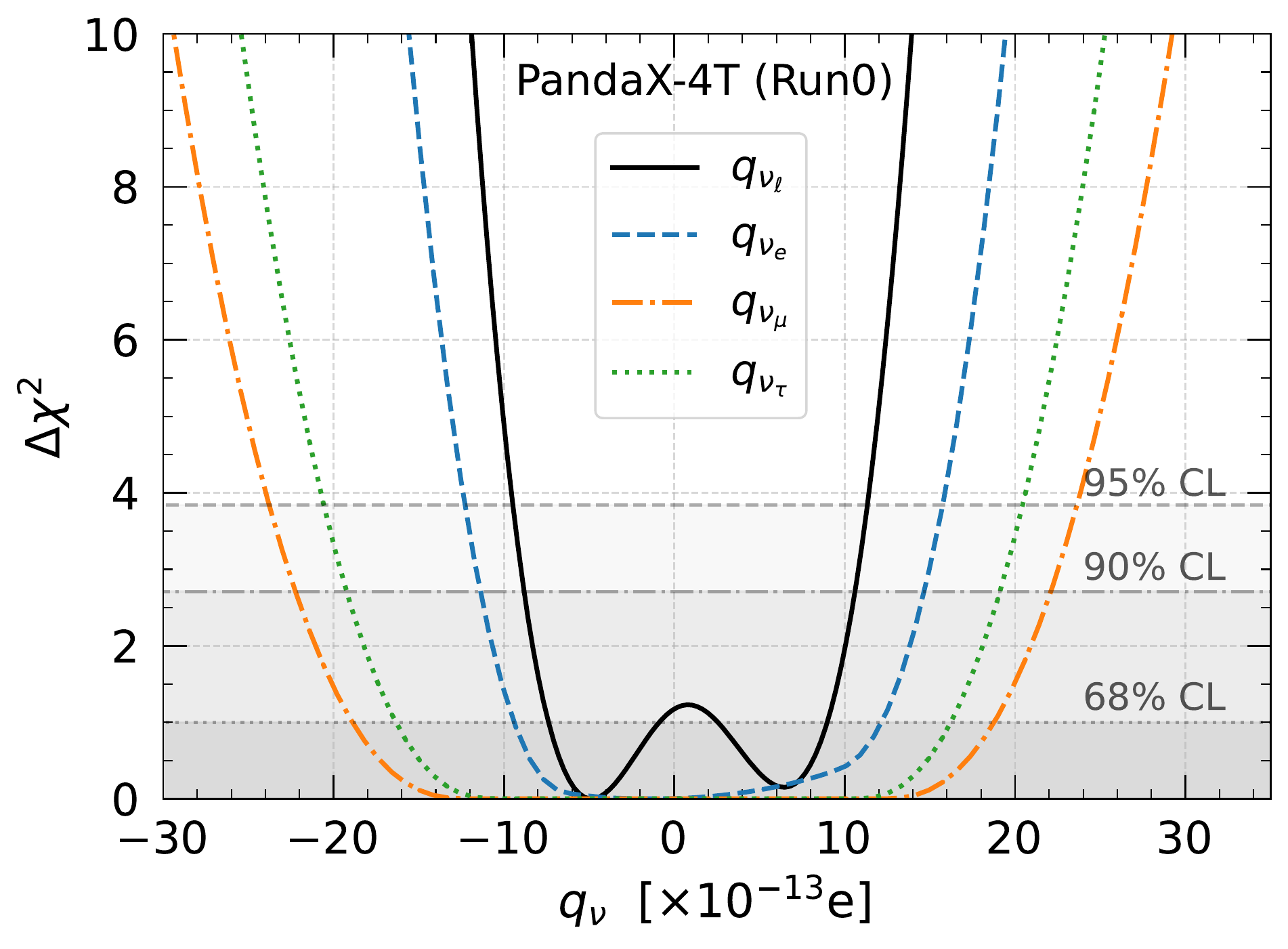}
	\includegraphics[scale=0.42]{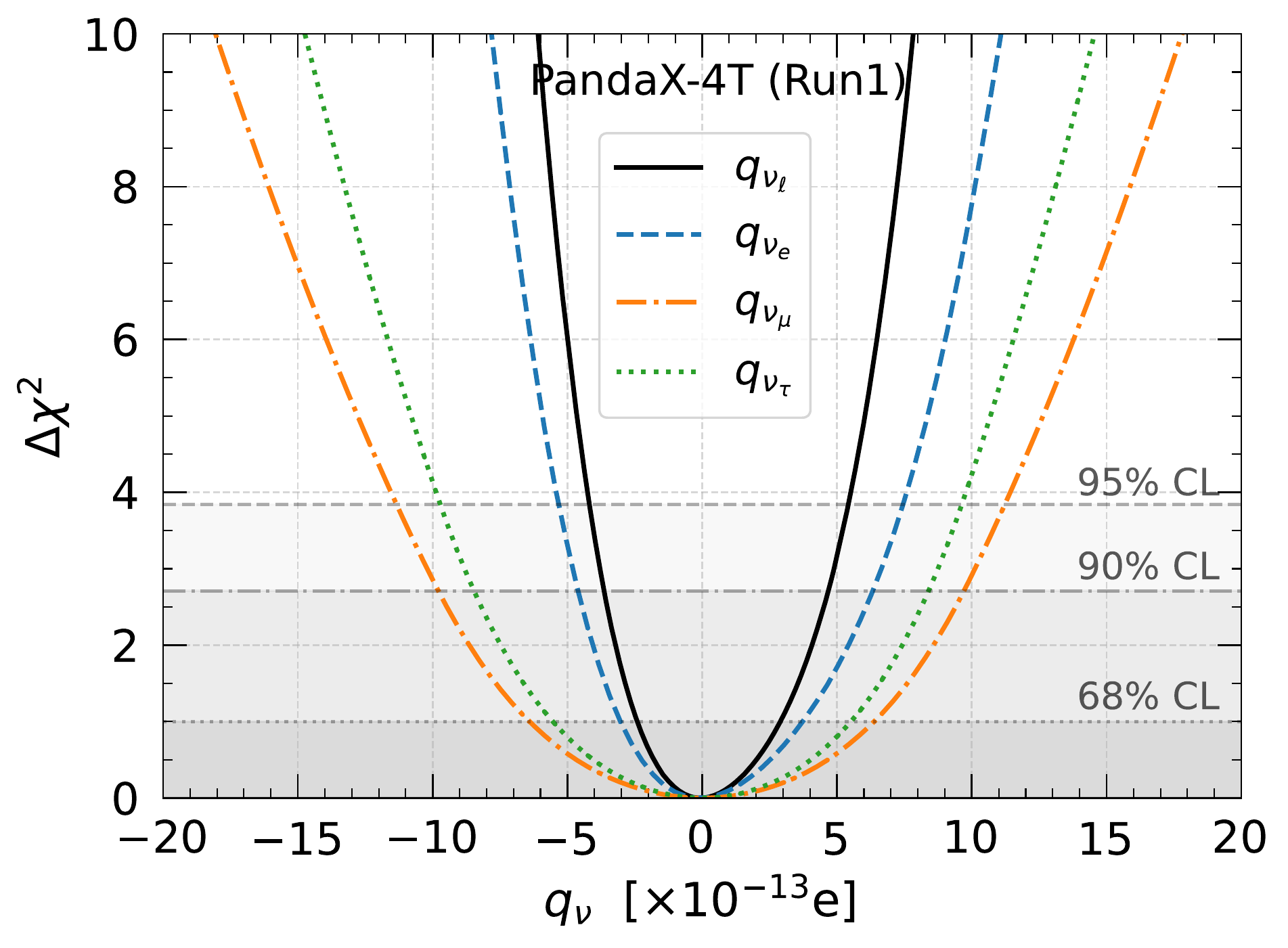}
	\\
	\hspace{9.6mm} (a) \hspace{76mm} (b)
	\\
	\includegraphics[scale=0.42]{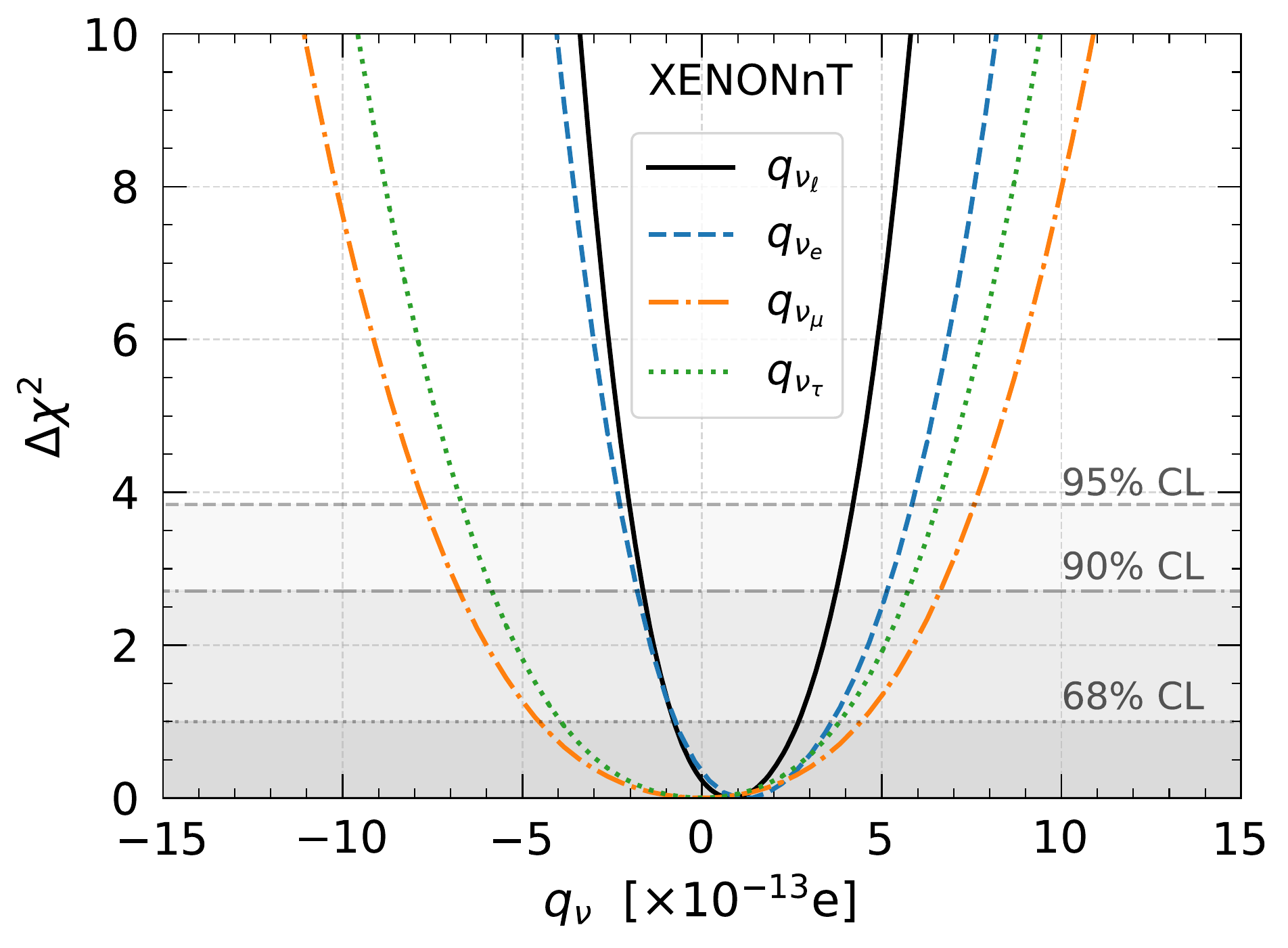}
    \\
	(c)
	\caption{One-dimensional $\Delta\chi^2$ distributions obtained from the analysis of recent data: (a) PandaX-4T Run0, (b) PandaX-4T Run1, and (c) XENONnT. The results are presented for the neutrino millicharge, including both the flavor-independent effective case and the flavor-dependent cases derived through marginalization over the three neutrino flavors.}
	\label{fig:nuMC}
\end{figure*}
\begin{figure*}[htb!]
	\centering
	\includegraphics[scale=0.31]{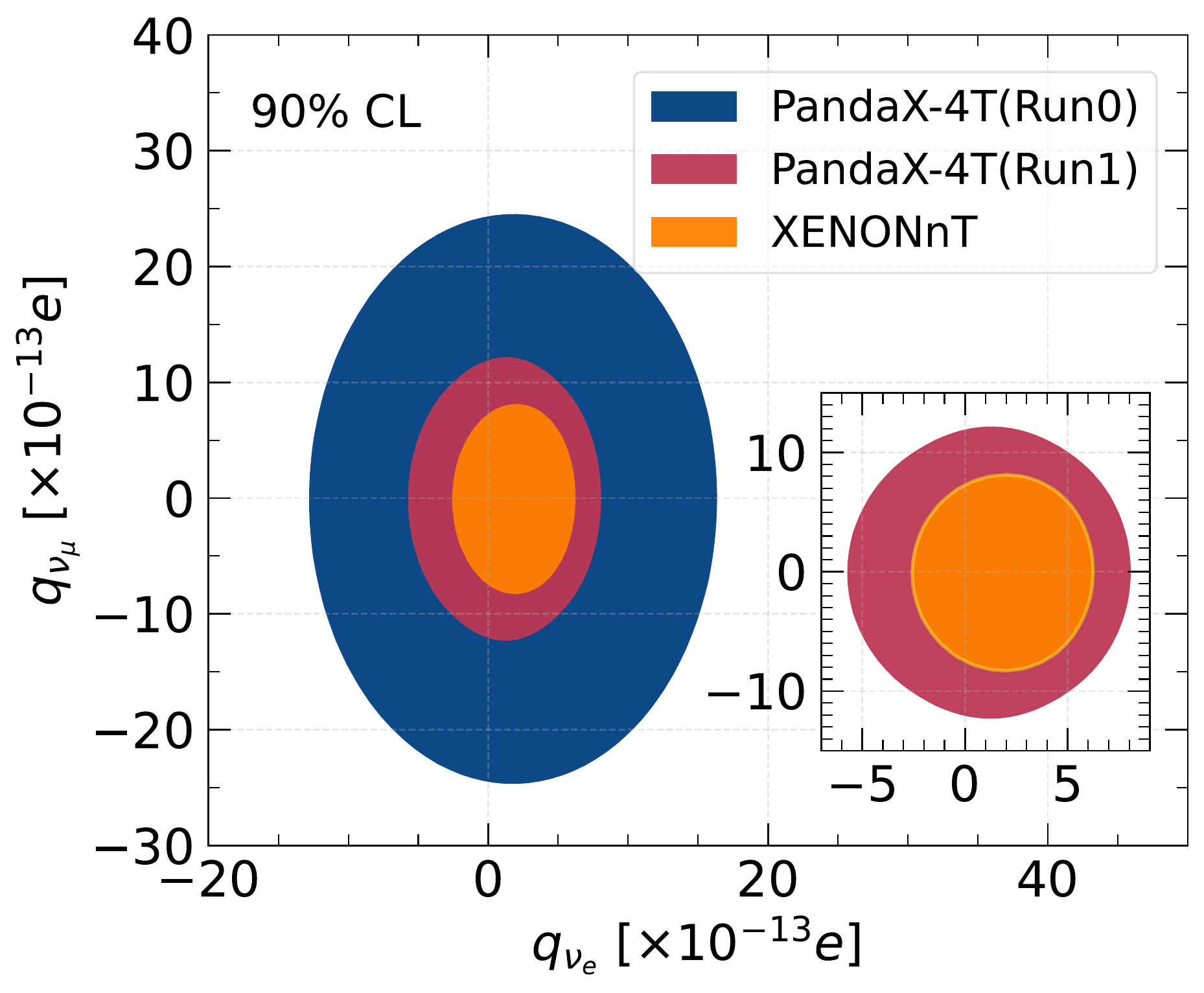}
	\includegraphics[scale=0.31]{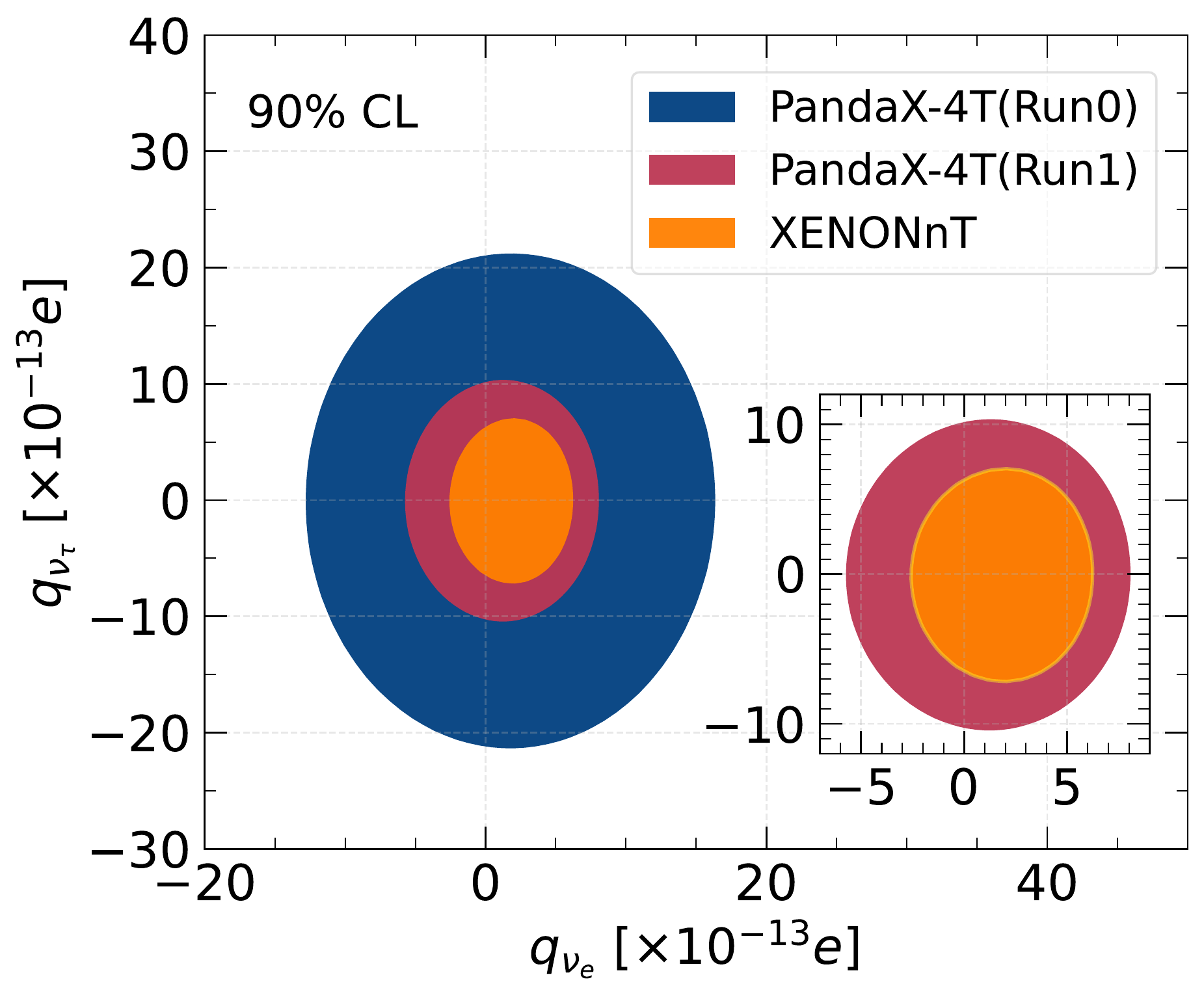}
    \includegraphics[scale=0.31]{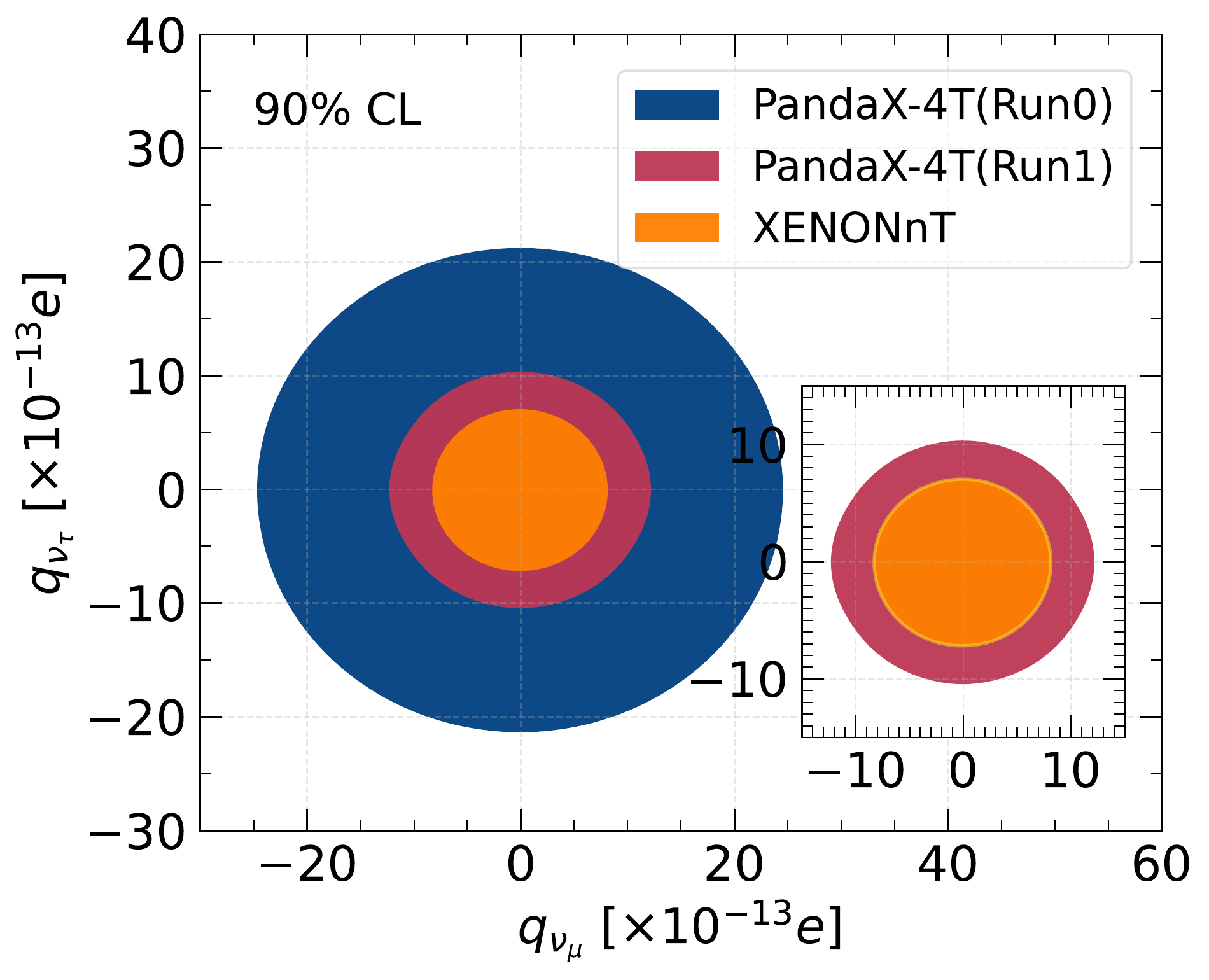}
	\\
	\hspace{7mm} (a) \hspace{53mm} (b) \hspace{53mm} (c)
	\vspace{-2mm}
       \caption{The $90\%$ CL allowed regions with 2 d.o.f. in the (a) $(q_{\nu_e},q_{\nu_\mu})$, (b) $(q_{\nu_e},q_{\nu_\tau})$, and (c) $(q_{\nu_\mu},q_{\nu_\tau})$ planes, obtained by marginalizing over the remaining neutrino flavor components. The results correspond to the analysis of PandaX-4T Run0 (blue region), PandaX-4T Run1 (red region), and XENONnT data (orange region).}
	\label{fig:nuMC_2dof}
\end{figure*}
\begin{figure*}[htb]
	\centering
	\includegraphics[scale=0.42]{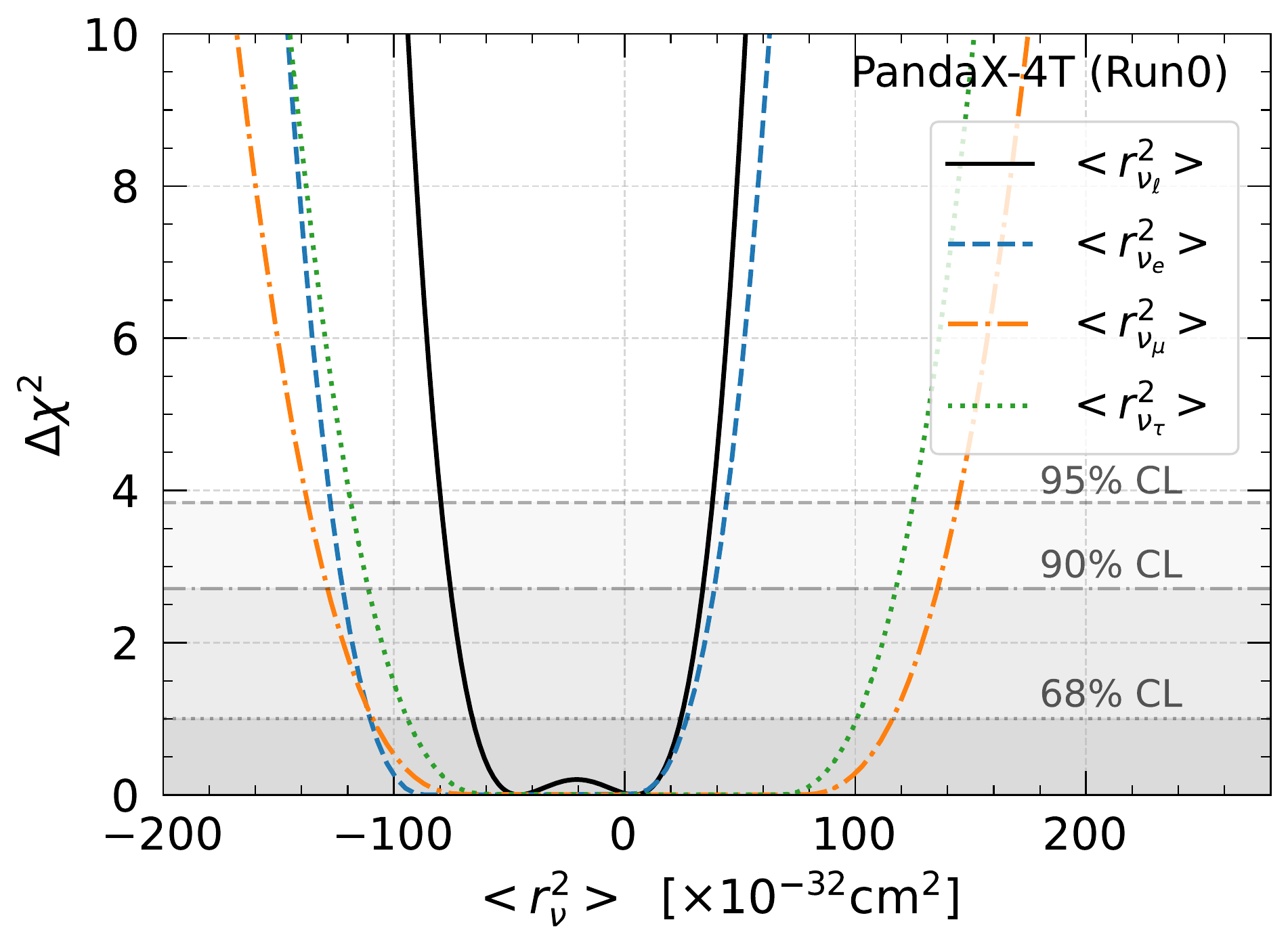}
	\includegraphics[scale=0.42]{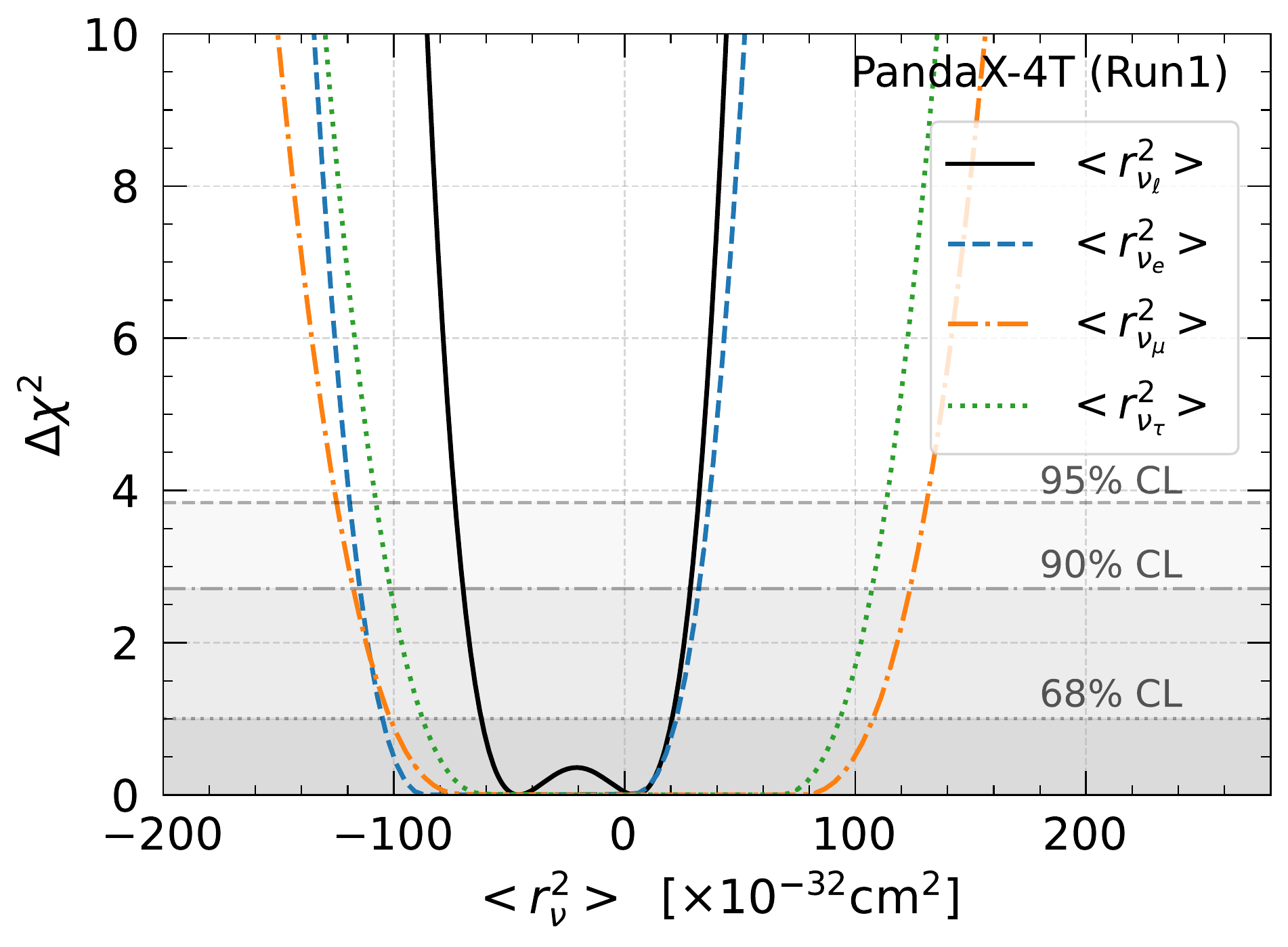}
	\\
	\hspace{9mm} (a) \hspace{76mm} (b)
	\\
	\includegraphics[scale=0.42]{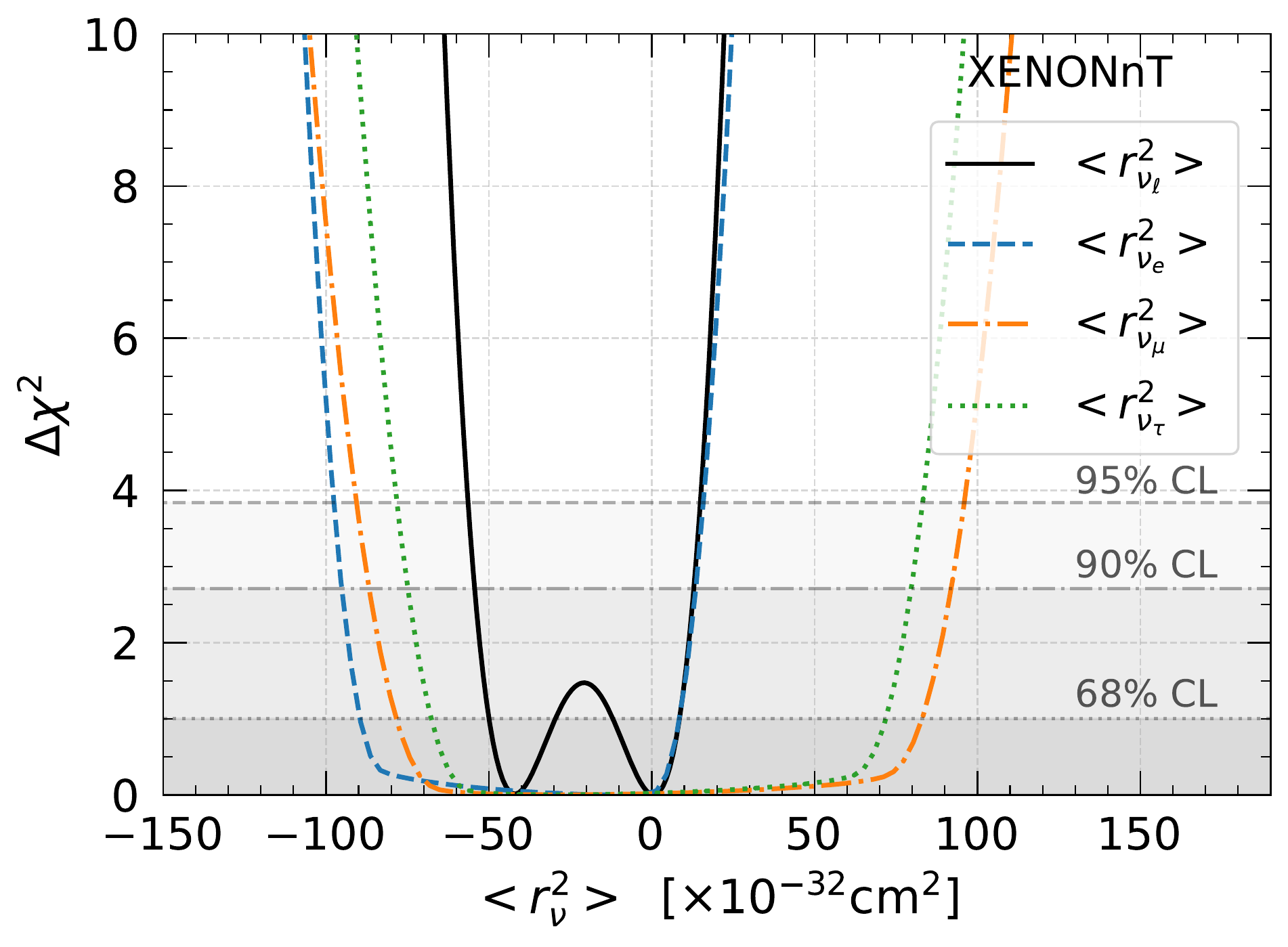}
    \\
	\hspace{9mm} (c)
	\vspace{-2mm}
	\caption{One-dimensional $\Delta\chi^2$ distributions obtained from the analysis of recent data: (a) PandaX-4T Run0, (b) PandaX-4T Run1, and (c) XENONnT. The results are presented for the neutrino charge radius, including both the flavor-independent effective case and the flavor-dependent cases derived through marginalization over the three neutrino flavors.}
	
	\label{fig:nuCR}
\end{figure*}
\begin{figure*}[htb!]
	\centering
	\includegraphics[scale=0.315]{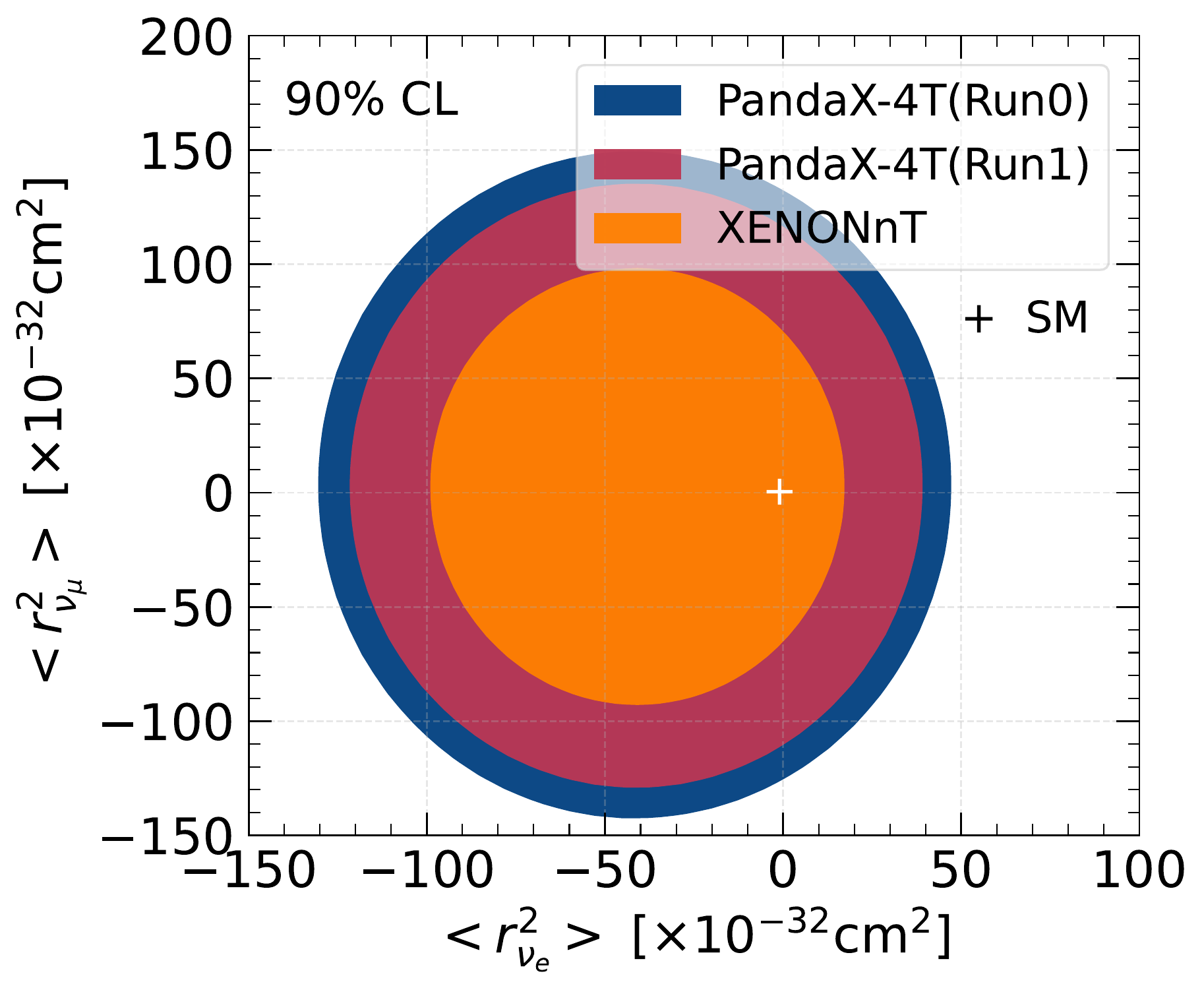}
	\includegraphics[scale=0.315]{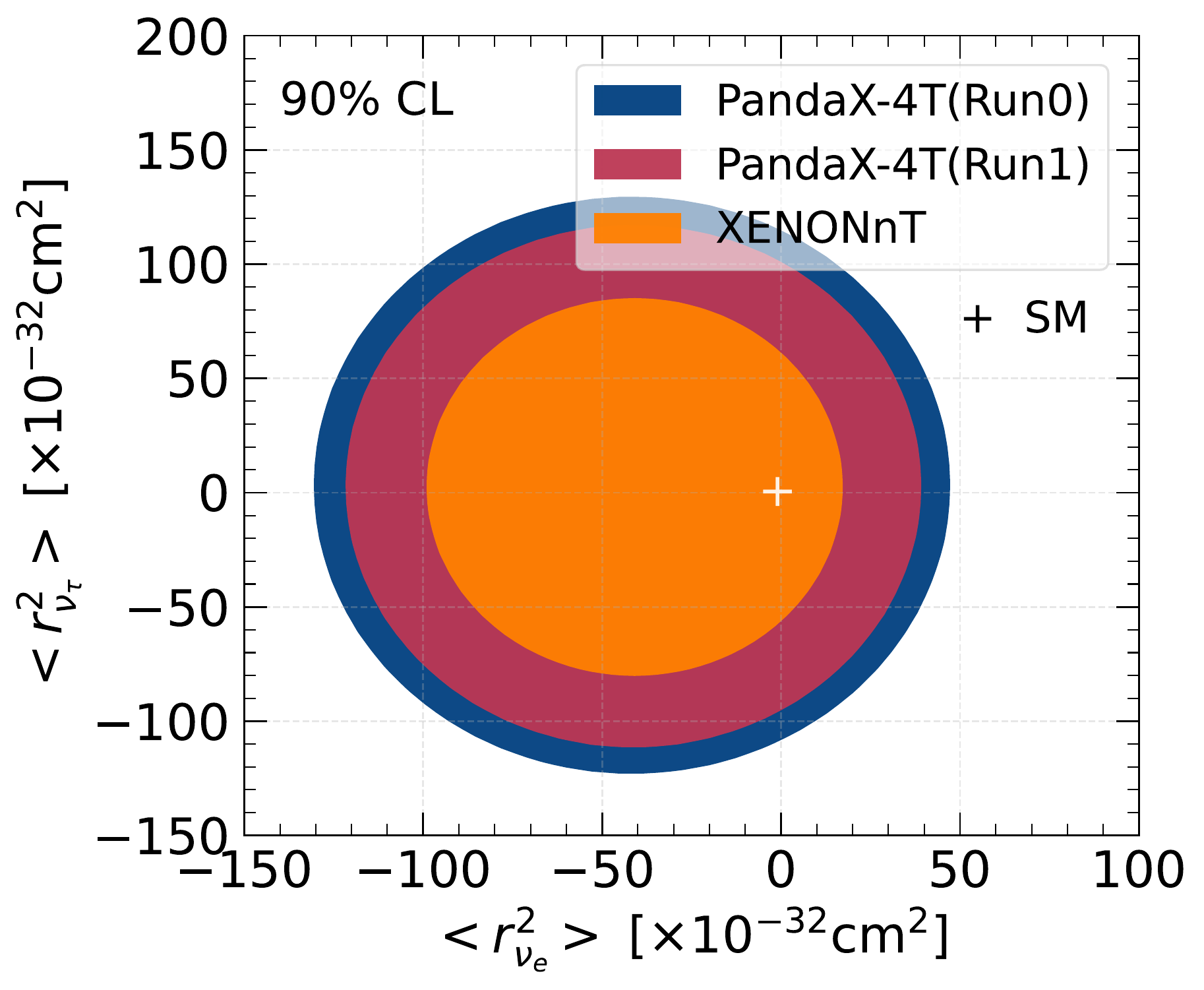}
    \includegraphics[scale=0.315]{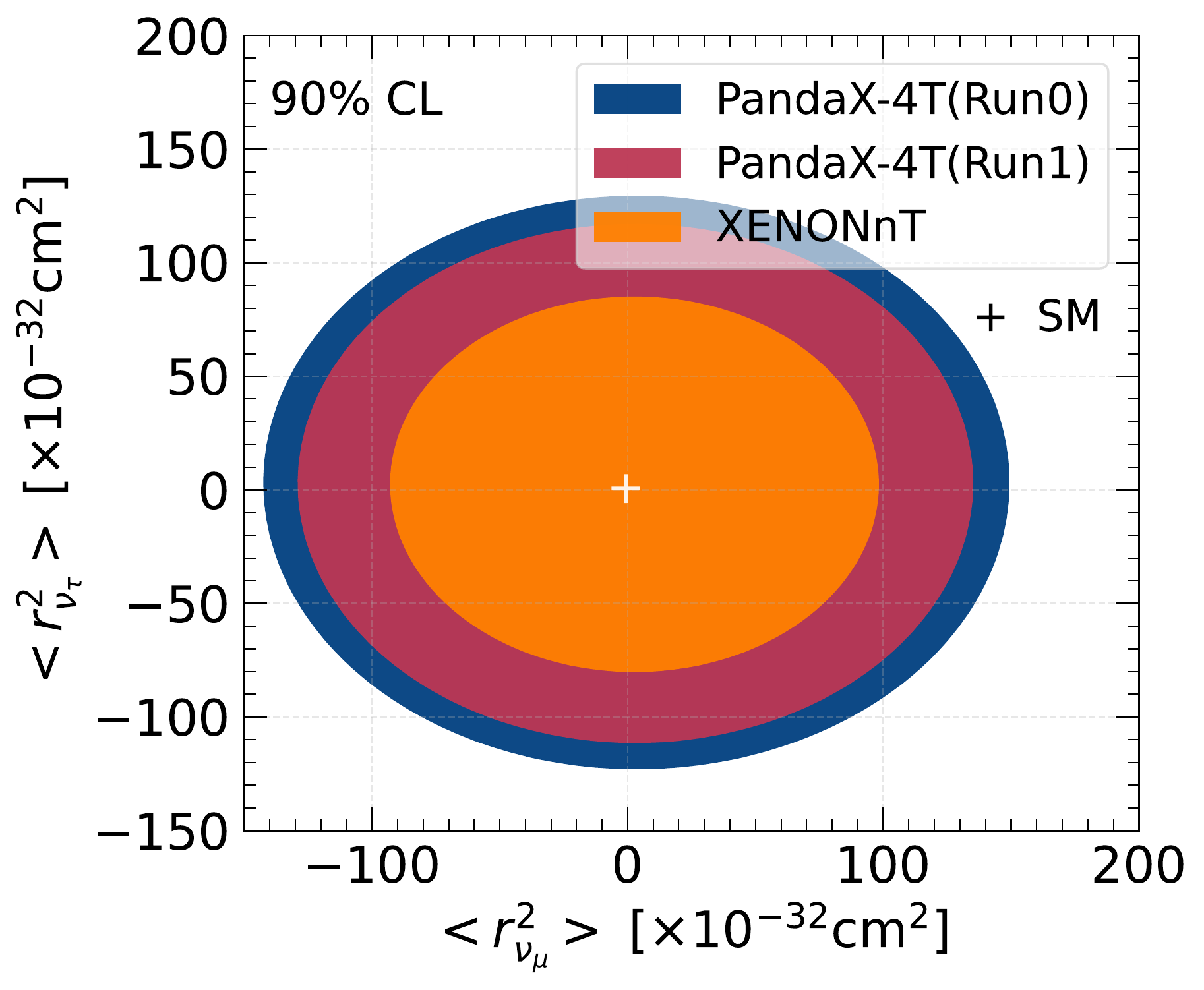}
	\\
	\hspace{7mm} (a) \hspace{53mm} (b) \hspace{53mm} (c)
	\vspace{-2mm}
     \\
	\caption{The $90\%$ CL allowed regions with 2 d.o.f. in the (a) $(\langle r^2_{\nu_e}\rangle,\langle r^2_{\nu_\mu}\rangle)$, (b)  $(\langle r^2_{\nu_e}\rangle,\langle r^2_{\nu_\tau}\rangle)$ and (c)  $(\langle r^2_{\nu_\mu} \rangle, \langle r^2_{\nu_\tau}\rangle)$  planes, obtained by marginalizing over the remaining neutrino flavor components.  
    The results correspond to the analysis of PandaX-4T-Run0 (blue region), PandaX-4T-Run1 (red region), and  XENONnT data (orange region).
	}
	\label{fig:nuCR_2dof}
\end{figure*}
Figure~\ref{fig:nuMM} presents the 1 d.o.f. $\Delta\chi^2$ profiles of the neutrino magnetic moment from analyzing PandaX-4T and XENONnT data. 
It is seen that for the flavor-independent case, the 90\% CL result of the XENONnT ($\mu_{\nu_\ell} \lesssim 0.72 \times 10^{-11} \mu_\text{B}$) is around three times stringent than ones of the PandaX-4T Run0 ($\mu_{\nu_\ell}  \lesssim 2.41 \times 10^{-11} \mu_\text{B}$) and two times stringent than the result of PandaX-4T Run1 ($\mu_{\nu_\ell}  \lesssim 1.37 \times 10^{-11} \mu_\text{B}$). 
Similar trends are followed by the flavor-dependent cases. 
In Figs.~\ref{fig:nuMM_2dof}(a), (b), and (c), we also provide the $90\%$ CL allowed regions with 2 d.o.f. in the 
$(\mu_{\nu_e},\mu_{\nu_\mu})$,  $(\mu_{\nu_e},\mu_{\nu_\tau})$ and $(\mu_{\nu_\mu},\mu_{\nu_\tau})$ planes, respectively. As can be seen in these figures, 
the XENONnT result provides the most stringent allowed region among the three datasets considered. Specifically, it improves upon the PandaX-4T Run0 allowed region by a factor of approximately 3, and is about 1.8 times tighter than the constraint obtained in PandaX-4T Run1. 
The shape and extent of the allowed regions reflect the interplay between experimental uncertainties and sensitivities across the different neutrino flavor combinations.

\begin{table*}[htbp]
	\centering
	\caption{Summary of the derived limits on the neutrino electromagnetic properties from PandaX-4T Run 0 and Run 1 datasets. 
		}
	\resizebox{\textwidth}{!}{%
		\begin{tabular}{lcccccc}
			\hline\hline
		\multirow{2}{1.5cm}{Property} 
			& \multicolumn{3}{c}{PandaX-4T (Run0)} 
			& \multicolumn{3}{c}{PandaX-4T (Run1)} \\ \cline{2-4} \cline{5-7}
			& 68\% CL & 90\% CL & 95\% CL & 68\% CL & 90\% CL & 95\% CL \\
			\hline
			$\mu_{\nu_\ell} \, (\times 10^{-11}\mu_B)$   & $\lesssim 2.04$ & $\lesssim 2.41$ & $\lesssim 2.57$ & $\lesssim 1.05$ & $\lesssim 1.37$ & $\lesssim 1.50$ \\
			$\mu_{\nu_e} \, (\times 10^{-11}\mu_B)$ & $\lesssim 2.29$ & $\lesssim 2.87$ & $\lesssim 3.13$ & $\lesssim 1.43$ & $\lesssim 1.87$ & $\lesssim 2.05$ \\
			$\mu_{\nu_\mu} \, (\times 10^{-11}\mu_B)$& $\lesssim 3.76$ & $\lesssim 4.73$ & $\lesssim 5.14$ & $\lesssim 2.36$ & $\lesssim 3.07$ & $\lesssim 3.37$ \\
			$\mu_{\nu_\tau} \, (\times 10^{-11}\mu_B)$& $\lesssim 3.25$ & $\lesssim 4.09$ & $\lesssim 4.45$ & $\lesssim 2.04$ & $\lesssim 2.66$ & $\lesssim 2.92$ \\
			\hline
			$q_{\nu_\ell} \, (\times 10^{-12} e)$   & $[-0.74,-0.09] \cup [0.26,0.89]$ & $[-0.88,1.06]$ & $[-0.95,1.13]$ & $[-0.24,0.29]$ & $[-0.36,0.47]$ & $[-0.42,0.54]$ \\
			$q_{\nu_e} \, (\times 10^{-12} e)$ & $[-0.94,1.21]$ & $[-1.14,1.47]$ & $[-1.23,1.58]$ & $[-0.30,0.37]$ & $[-0.46,0.63]$ & $[-0.53,0.74]$ \\
			$q_{\nu_\mu} \, (\times 10^{-12} e)$& $[-1.89,1.87]$ & $[-2.22,2.21]$ & $[-2.38,2.36]$ & $[-0.64,0.64]$ & $[-0.98,0.97]$ & $[-1.13,1.12]$ \\
			$q_{\nu_\tau} \, (\times 10^{-12} e)$& $[-1.63,1.62]$ & $[-1.92,1.91]$ & $[-2.06,2.04]$ & $[-0.55,0.55]$ & $[-0.85,0.84]$ & $[-0.97,0.96]$ \\
			\hline
			$\langle r_{\nu_\ell}^2 \rangle \, (\times 10^{-32}\,\text{cm}^2)$   & $[-65.94,24.40]$ & $[-75.44,33.92]$ & $[-79.61,38.08]$ & $[-62.10, 20.61]$ & $[-70.01,28.52]$ & $[-73.49, 31.99]$ \\
			$\langle r_{\nu_e}^2 \rangle \, (\times 10^{-32}\,\text{cm}^2)$ & $[-110.30,26.87]$ & $[-122.14,38.85]$ & $[-127.52,44.12]$ & $[-104.88, 22.41]$ & $[-114.71,32.13]$ & $[-119.18,36.54]$ \\
			$\langle r_{\nu_\mu}^2 \rangle \, (\times 10^{-32}\,\text{cm}^2)$& $[-109.11,116.09]$ & $[-128.73,135.68]$ & $[-137.44,144.34]$ & $[-101.50,107.67]$ & $[-117.60,123.70]$ & $[-124.81,130.90]$ \\
			$\langle r_{\nu_\tau}^2 \rangle \, (\times 10^{-32}\,\text{cm}^2)$& $[-94.12,100.68]$ & $[-111.13,117.58]$ & $[-118.64,125.10]$ & $[-87.58,93.31]$ & $[-101.48,107.23]$ & $[-107.73,113.41]$ \\
			\hline\hline
		\end{tabular}%
	}	\label{tab:rslt_pndx}
\end{table*}
For the neutrino millicharge, the 1 d.o.f. $\Delta\chi^2$ profiles from the analyses of the PandaX-4T and XENONnT data are shown in Fig.~\ref{fig:nuMC}. The 90\% CL limits of the flavor-independent case indicate that the XENONnT limit of $-1.64 \times10^{-13} e \lesssim q_{\nu_\ell}\lesssim 3.73 \times10^{-13} e$  is a few times more stringent than both the PandaX-4T (Run0) limit of $-8.81 \times10^{-13} e \lesssim q_{\nu_\ell}\lesssim 10.6 \times10^{-13} e$ and the PandaX-4T (Run1) limit of $-3.64 \times10^{-13} e \lesssim q_{\nu_\ell}\lesssim 4.68 \times10^{-13} e$. The $90\%$ CL allowed regions with 2 d.o.f. in the 
$(q_{\nu_e},q_{\nu_\mu})$, $(q_{\nu_e},q_{\nu_\tau})$ and $(q_{\nu_\mu},q_{\nu_\tau})$  
planes are shown in Figs.~\ref{fig:nuMC_2dof}(a), (b) and (c). 
As clearly seen, the allowed regions derived from PandaX-4T Run0 are significantly broader in all three parameter spaces, indicating weaker constraints. In contrast, the XENONnT provides substantially tighter allowed regions, reflecting a notable improvement in sensitivity. The  PandaX-4T Run1 results are comparable to those of XENONnT, although slightly weaker in most of the parameter spaces.
The insets in each panel effectively illustrate the comparative proximity of the central allowed regions for PandaX-4T Run1 and XENONnT, both of which are orders of magnitude tighter than the original PandaX-4T Run0 limits.
\begin{table*}[htbp]
	\centering
	\caption{Summary of our derived limits on the neutrino electromagnetic properties from XENONnT data.}
		\begin{tabular}{lccc}
			\hline\hline
			\multirow{2}{1.5cm}{Property}  & \multicolumn{3}{c}{XENONnT} \\ \cline{2-4}
			& 68\% CL & 90\% CL & 95\% CL \\
			\hline
			$\mu_{\nu_\ell} \, (\times 10^{-11}\mu_B)$   & $\lesssim 0.52$ & $\lesssim 0.72$ & $\lesssim 0.82$ \\
			$\mu_{\nu_e} \, (\times 10^{-11}\mu_B)$ & $\lesssim 0.69$ & $\lesssim 0.98$ & $\lesssim 1.11$ \\
			$\mu_{\nu_\mu} \, (\times 10^{-11}\mu_B)$& $\lesssim 1.16$ & $\lesssim 1.62$ & $\lesssim 1.83$ \\
			$\mu_{\nu_\tau} \, (\times 10^{-11}\mu_B)$& $\lesssim 1.00$ & $\lesssim 1.40$ & $\lesssim 1.58$ \\
			\hline
			$q_{\nu_\ell} \, (\times 10^{-12} e)$   & $[-0.08,0.27]$ & $[-0.16,0.37] $ & $[-0.20,0.42]$ \\
			$q_{\nu_e} \, (\times 10^{-12} e)$ & $[-0.07,0.36]$ & $[-0.18,0.51]$ & $[-0.23,0.58]$ \\
			$q_{\nu_\mu} \, (\times 10^{-12} e)$& $[-0.45,0.44]$ & $[-0.68,0.66]$ & $[-0.77,0.76]$ \\
			$q_{\nu_\tau} \, (\times 10^{-12} e)$& $[-0.39,0.38]$ & $[-0.59,0.57]$ & $[-0.67,0.65]$ \\
			\hline
			$\langle r_{\nu_\ell}^2 \rangle \, (\times 10^{-32}\,\text{cm}^2)$   & $[-49.95,-29.61] \cup [-11.94,8.52]$ & $[-54.36, 12.90]$ & $[-56.38,14.91]$ \\
			$\langle r_{\nu_e}^2 \rangle \, (\times 10^{-32}\,\text{cm}^2)$ & $[-89.56,8.42]$ & $[-95.15,13.53]$ & $[-97.55,15.82]$ \\
			$\langle r_{\nu_\mu}^2 \rangle \, (\times 10^{-32}\,\text{cm}^2)$& $[-78.36,82.98]$ & $[-86.75,91.90]$ & $[-90.62,95.94]$ \\
			$\langle r_{\nu_\tau}^2 \rangle \, (\times 10^{-32}\,\text{cm}^2)$& $[-67.65,71.84]$ & $[-74.85,79.63]$ & $[-78.20,83.17]$ \\
			\hline\hline
		\end{tabular}%
			\label{tab:rslt_xenon}
\end{table*}

Finally, Figure~\ref{fig:nuCR} shows the 1 d.o.f. $\Delta\chi^2$ profiles of the neutrino charge radius from our analyses of PandaX-4T and XENONnT data. For the flavour-independent results at 90\% CL, the PandaX-4T (Run1) limit ($-70.01 \times 10^{-32} \text{ cm}^2 \lesssim \langle r_{\nu_\ell}^2\rangle \lesssim 28.52 \times 10^{-32} \text{cm}^2$) is moderately improved compared to PandaX-4T (Run0) ($-75.44 \times 10^{-32} \text{ cm}^2 \lesssim \langle r_{\nu_\ell}^2\rangle \lesssim 33.92 \times 10^{-32} \text{ cm}^2$), reflecting the increased exposure. When compared to the XENONnT bound ($-54.36 \times 10^{-32} \text{ cm}^2 \lesssim \langle r_{\nu_\ell}^2\rangle \lesssim 12.90 \times 10^{-32} \text{ cm}^2 $), however, the PandaX-4T (Run1) constraint is somewhat weaker, as the XENONnT interval is notably narrower and thus provides the most stringent limit among the three.

In Figs.~\ref{fig:nuCR_2dof}(a), (b) and (c), the $90\%$ CL allowed regions with 2 d.o.f. are shown in the
$(\langle r^2_{\nu_e} \rangle,\langle r^2_{\nu_\mu} \rangle)$,  $(\langle r^2_{\nu_e}\rangle,\langle r^2_{\nu_\tau} \rangle)$ and  $(\langle r^2_{\nu_\mu}\rangle,\langle r^2_{\nu_\tau}\rangle)$
planes, respectively. The SM expectation, derived from one-loop radiative corrections as shown in Eq. \eqref{eq:SMvalues}, is denoted by the white cross.
As can be seen in these figures, again, the XENONnT bound is stronger than the one from PandaX-4T, while the PandaX-4T Run0 is slightly weaker than that from PandaX-4T Run1. Namely, the XENONnT experiment provides the leading constraints, while PandaX-4T Run1 exhibits marked improvements over its earlier Run0, closing the gap with XENONnT and promising even tighter bounds with future datasets.

\begin{table*}[ht]
	\caption{
		90\% CL limits on the neutrino magnetic moment, millicharge, and charge radius from this work and previous studies.
    }
	\begin{center}
		\begin{tabular}{ c c l c l c l }
			\hline
			\hline
			Flavor & & $\mu_\nu (\times 10^{-11}\mu_\mathrm{B})$ & & $q_\nu (\times 10^{-12}e)$ & & $\langle r_\nu^2 \rangle (\times 10^{-32}\mathrm{cm}^2)$
            \\
			\hline
            \multirow{3}{1em}{$\nu_\ell$} & & $\lesssim 2.41$ (\textbf{PandaX-4T Run0}) & & $[-0.88,1.06]$ (\textbf{PandaX-4T Run0}) & & $[-75.44,33.92]$ (\textbf{PandaX-4T Run0}) 
            \\
            & & $\lesssim 1.37$ (\textbf{PandaX-4T Run1}) & & $[-0.36,0.47]$ (\textbf{PandaX-4T Run1}) & & $[-70.01,28.52]$ (\textbf{PandaX-4T Run1}) 
            \\
             & & $\lesssim0.72$ (\textbf{XENONnT}) & & $[-0.16,0.37]$ (\textbf{XENONnT}) & & $[-54.36, 12.90]$ (\textbf{XENONnT})
            \\
            & & $\lesssim1.1$ ({LZ}) \cite{AtzoriCorona:2022jeb} & & $[-0.15,0.15]$ ({LZ}) \cite{AtzoriCorona:2022jeb} & & -
            \\
            & & $\lesssim 1.8$ ({XMASS-I}) \cite{XMASS:2020zke} & & $\lesssim 5.4$ ({XMASS-I} \cite{XMASS:2020zke}) & & -
            \\
            & & $\lesssim2.8$ ({BOREXINO}) \cite{Borexino:2017fbd} & & - & & -
            \\
             \hline
            \multirow{8}{1em}{$\nu_e$} 
            & & $\lesssim 1.5$ (LZ) \cite{AtzoriCorona:2022jeb} & & $[-0.21,0.2]$ (LZ) \cite{AtzoriCorona:2022jeb} & & $[-110.4,26.4]$ (LZ) \cite{Giunti:2023yha} 
            \\
            & & $\lesssim 3.9$ (BOREXINO) \cite{Borexino:2017fbd} & & $\lesssim 7.3$ (XMASS-I) \cite{XMASS:2020zke}  & & $[-5.94,8.28]$ (LSND) \cite{LSND:2001akn}
            \\
            & & $\lesssim 7.4$ (TEXONO) \cite{TEXONO:2006xds} & & $[-1.0,1.0]$ (TEXONO) \cite{Chen:2014dsa} & & $[-4.2,6.6]$ (TEXONO) \cite{TEXONO:2009knm}
            \\
            & & $\lesssim 2.9$ (GEMMA) \cite{Beda:2012zz} & & $[-0.6,0.6]$ (CONUS+) \cite{AtzoriCorona:2025ygn} & & $[-76.0,-57.0] \cup [-8.0,11.0]$ (CONUS+) \cite{AtzoriCorona:2025ygn}
            \\
            & & $\lesssim 12.0$ (CONUS+) \cite{AtzoriCorona:2025ygn} & & $[-9.3,9.5]$ (Dresden-II) \cite{AtzoriCorona:2022qrf} & & $[-56.7,40.8]$ (Dresden-II) \cite{AtzoriCorona:2022qrf}
            \\
            && $\lesssim 21.3$ (Dresden-II) \cite{AtzoriCorona:2022qrf} && $[-260.0,260.0]$ (COHERENT) \cite{DeRomeri:2022twg} && $[-61.2,-48.2]\cup [-4.7,2.2]$ (COHERENT) \cite{DeRomeri:2022twg}
            \\
            && $\lesssim 360.0$ (COHERENT) \cite{DeRomeri:2022twg}  && - && -
            \\
            \hline
            \multirow{3}{1em}{$\nu_\mu$} 
            & & $\lesssim2.3$ (LZ) \cite{AtzoriCorona:2022jeb} & & $[-0.31,0.31]$ (LZ) \cite{AtzoriCorona:2022jeb} & & $[-101.8,105.5]$ (LZ) \cite{Giunti:2023yha}
            \\
            & & $\lesssim5.8$ (BOREXINO) \cite{Borexino:2017fbd} & & $\lesssim 11.0$ (XMASS-I) \cite{XMASS:2020zke} & & $[-1.2.1.2]$ (CHARM-II) \cite{Hirsch:2002uv}
            \\
            & & $\lesssim 240.0$ (COHERENT) \cite{DeRomeri:2022twg} && $[-140.0,140.0]$ (COHERENT) \cite{DeRomeri:2022twg} && $[-58.2,-52.1]$ (COHERENT) \cite{DeRomeri:2022twg}
            \\
            \hline
            \multirow{2}{1em}{$\nu_\tau$} 
            & & $\lesssim2.1$ (LZ) \cite{AtzoriCorona:2022jeb} & & $[-0.28,0.28]$ (LZ) \cite{AtzoriCorona:2022jeb} & & $[-101.8,105.5]$ (LZ) \cite{Giunti:2023yha}
            \\
            & & $\lesssim5.8$ (BOREXINO) \cite{Borexino:2017fbd} & & $\lesssim 11.0$ (XMASS-I) \cite{XMASS:2020zke} & & -
            \\
            \hline
            \hline
		\end{tabular}
	\end{center}
	\label{tab:comparison}
\end{table*}

We provide the summary of our results for 1 d.o.f. in Table~\ref{tab:rslt_pndx} for PandaX-4T Run0 and Run1 datasets, and Table~\ref{tab:rslt_xenon} for XENONnT data. We give 68\% CL, 90\% CL and 95\% CL limits for each neutrino's electromagnetic property. These values can be read directly from Fig.~\ref{fig:nuMM} for neutrino magnetic moment, Fig.~\ref{fig:nuMC} for neutrino milli charge, and Fig.~\ref{fig:nuCR} for neutrino charge radius. 

The limits on each neutrino electromagnetic property for our $\nu_\ell$ case at 90\% CL in comparison with previous studies are presented in Table~\ref{tab:comparison}. 
Alongside results from direct detection experiments, we also include limits derived from nuclear reactors, spallation neutron sources, and other relevant facilities to provide a complementary overview.
It should be noted that the comparison of different neutrino flavors is possible with the caveat that they may have been obtained with different values of neutrino survival probability and the neutrino oscillation parameters, as well as different treatments of the systematic uncertainties from each experiment. Consequently, we present the comparison qualitatively, by comparing our effective flavor-independent results (corresponding to our $\nu_\ell$ results) with flavor-specific results from other experiments. 
In general, we obtain improved and/or competitive results in this study for each electromagnetic quantity.

For the neutrino magnetic moment $\mu_{\nu_\ell} $, the XENONnT analysis has achieved the most stringent constraint. It represents a significant improvement over PandaX-4T Run0 and Run1 and outperforms other direct detection experiments of LZ, and XMASS-I, as well as solar neutrino experiment of BOREXINO. 
In addition, these are more sensitive than limits obtained from reactor experiments of TEXONO, GEMMA, CONUS+, and Dresden-II as we can see in the flavor-dependent cases. To complement, results from the COHERENT experiment is also shown. 

Regarding the neutrino millicharge $q_{\nu_\ell}$, both PandaX-4T Run1 and XENONnT analysis yields tightest limits. The PandaX-4T Run1 result shows a clear improvement over Run0, reducing the previously allowed region by nearly a factor of two. Among the direct detection results, the obtained limits are yet to reach the LZ, while surpassing the XMASS-I result.  Furthermore, the obtained bounds are more stringent than those derived from reactor-based experiments such as TEXONO, CONUS+, and Dresden-II, as well as from the pion-stopped experiment at COHERENT.

Moving on to the case of the neutrino charge radius $\langle r_{\nu_\ell}^2 \rangle$, the limit from PandaX-4T Run1 is noticeably improved compared to Run0, reflecting the increased exposure. Among the direct detection datasets, XENONnT provides the most stringent constraint, yielding a narrower allowed interval than PandaX-4T. Our results thus indicate that the PandaX-4T bounds are competitive yet slightly weaker than those from XENONnT, while both are comparable to the sensitivity of LZ. Nevertheless, none of the direct detection limits have yet reached the precision achieved by accelerator-based measurements such as LSND and CHARM-II. Still, these bounds surpass those obtained from reactor experiments including TEXONO, CONUS+, and Dresden-II, as well as from the pion-stopped COHERENT experiment.

The improved sensitivity in PandaX-4T Run1 over Run0 highlights enhanced background suppression and systematic control within the experiment. Regarding the XENONnT, the data provides a highly competitive constraint. These further consolidating the reliability and consistency of direct detection results across different experimental setups. Overall, these findings demonstrate the capability of modern direct detection experiments to probe neutrino electromagnetic properties with unprecedented precision, yielding more stringent constraints than several dedicated neutrino experiments.

We emphasize that our limits are consistent with the literature.
Our findings confirm previous results from a similar analysis of XENONnT data conducted in Ref. \cite{Giunti:2023yha}. 
Additionally, other phenomenological studies, including Ref.\cite{A:2022acy} and Ref.\cite{Khan:2023b},  have adopted modest approaches in the treatment of uncertainties, with the latter utilizing data up to 160 keV electron recoil from XENONnT.
They reported upper limits on the neutrino magnetic moment at the  90 \% CL as $|\mu_{\nu_e}| \lesssim 11.5 \times 10^{-12} \mu_\text{B}$ \cite{Giunti:2023yha}, \(|\mu_{\nu_e}| \lesssim 13.9 \times 10^{-12} \mu_\text{B}\) \cite{A:2022acy}, and $|\mu_{\nu_e}| \lesssim 8.5 \times 10^{-12} \mu_\text{B}$ \cite{Khan:2023b}.
%
Our analysis results in an upper limit of $|\mu_{\nu_e}| \lesssim 9.8 \times 10^{-12} \mu_\text{B}$, showing good agreement with the magnitude of previously published limits.
In the case of other neutrino electromagnetic properties, notable improvements, by factors of a few, have been observed.
We anticipate that this behavior is due to the different choice of oscillation probabilities and parameters assumed. 
Moreover, we find that the predicted event rates of the solar neutrino background lie within the best-fit values reported in the experiments, as we mentioned in the previous section. 

Apart from these limits, we would like to mention that there are also limits that exist from cosmological sources. For instance, Big-Bang Nucleosynthesis (BBN) sets a limit on the magnetic moment of $\mu_\nu\lesssim 4.0 \times 10^{-12} \mu_{\text{B}}$ \cite{Grohs:2023xwa}, which is a few times less stringent than the limits obtained in our work. Another notable constraint arises from the analysis of the supernova SN1987A data, yielding a minimum upper limit of $\mu_\nu\lesssim 0.7\times 10^{-12} \mu_{\text{B}}$ \cite{Kuznetsov:2009zm}, a result that is comparable to our findings. In the case of the neutrino millicharge, $q_\nu$, the limits inferred from SN1987A and solar cooling considerations are more stringent than those obtained here, with values around $2.0\times 10^{-15} e - 2.0 \times 10^{-17} e$ \cite{Barbiellini:1987zz} and $q_\nu \lesssim 6.0 \times 10^{-14} e$ \cite{Raffelt:1999gv}. As for the charge radius, $\langle r_\nu^2 \rangle$, the bounds from BBN, $|\langle r_\nu^2 \rangle|\lesssim 0.7 \times 10^{-32} \text{ cm}^2$ \cite{Grifols:1986ed}, and from SN1987A, $|\langle r_\nu^2 \rangle|\lesssim 0.2 \times 10^{-32} \text{ cm}^2$ \cite{Altherr:1993hb}, remain among the most stringent limits currently available on this quantity.

Overall, our analysis represents a significant advance in the direct detection limits on neutrino electromagnetic properties. The consistent improvement from PandaX-4T Run0 to Run1, together with the most stringent constraint provided by XENONnT, highlights the importance of large-scale, low-background experiments in probing neutrino physics beyond the SM. These findings not only place tighter bounds on various new physics scenarios but also provide valuable benchmarks for next-generation detectors and future phenomenological studies.

\section{Summary and Conclusions}\label{sec:summ}
We studied neutrino electromagnetic properties, including the neutrino magnetic moment, millicharge, and charge radius, using the process of solar neutrino-electron scattering at direct detection experiments. 
We put limits on these properties by analyzing recent low-energy data from direct detection experiments.

We presented event rate predictions of neutrino-electron scattering for each neutrino electromagnetic property using solar neutrino fluxes.  
In general, we observe that these neutrino electromagnetic properties can be manifested as deviations from the SM spectrum, especially in the low-energy region. This indicates the need for enhancing detector sensitivity in this domain.

The new limits on the neutrino electromagnetic properties we deduced came from the most recent results from PandaX-4T Run0, PandaX-4T Run1, and XENONnT. These results have been compared with the limits obtained from other experiments. 
Adopting a more conservative treatment of systematic uncertainties, we achieve limits that are up to a few times tighter than those obtained in earlier analyses utilizing XENONnT data, while also delivering competitive and robust limits from the PandaX-4T Run0 and Run1 datasets.

To conclude, we have demonstrated that neutrino-electron scattering in direct detection experiments can be used to probe neutrino electromagnetic properties to obtain competitive constraints. It indicates that this feature can be implemented to search for other BSM physics. 

\section{Acknowledgments}
This work was supported by the Scientific and Technological Research Council of Türkiye (TUBITAK) under the project no: 124F416. The work of A.B.B. was also supported in part by the U.S. National Science Foundation Grant No. PHY-2411495.

\end{document}